\documentclass[prl,twocolumn,showpacs,amsmath,amssymb,superscriptaddress,longbibliography]{revtex4-1}
\usepackage{psfrag,graphicx}
\usepackage{amsfonts,amssymb,amsmath}      
\usepackage{graphicx}
\usepackage{dcolumn}
\usepackage{bm}
\usepackage{float}
\usepackage{hyperref}
\usepackage{accents}
\usepackage{bbding}
\usepackage{color,soul}
\usepackage{epstopdf}
\usepackage{changepage}
\usepackage[normalem]{ulem}
\usepackage{dsfont}

\newcommand{\erf}[1]{Eq.~(\ref{#1})}
\newcommand{\beq}{\begin{equation}}
\newcommand{\eeq}{\end{equation}}

\newcommand{\dg}{^\dagger}

\newcommand{\smallfrac}[2]{\mbox{$\frac{#1}{#2}$}}

\newcommand{\half}{\smallfrac{1}{2}}
\newcommand{\bra}[1]{\langle{#1}|}
\newcommand{\ket}[1]{|{#1}\rangle}
\newcommand{\ip}[2]{\langle{#1}|{#2}\rangle}
\newcommand{\op}[2]{\ket{#1}\bra{#2}}

\newcommand{\Tr}{\text{Tr}}

\newcommand{\s}[1]{\hat{\sigma}_{#1}}

\newcommand{\ex}[1]{\langle{#1}\rangle}
\newcommand{\dd}{{\rm d}}

\newcommand{\ie}{{\em i.e.}}

\newcommand{\past}[1]{\overleftarrow{#1}}
\newcommand{\fut}[1]{\overrightarrow{#1}}
\newcommand{\both}[1]{\overleftrightarrow{#1}}
\newcommand{\fil}{_{\text F}}
\newcommand{\rfil}{_{\text R}}
\newcommand{\sm}{_{\text S}}
\newcommand{\swv}{_{\rm SWV}}
\newcommand{\god}{_{\text T}}
\newcommand{\inv}{^{-1}}
\newcommand{\bx}{{\bf x}}

\newcommand{\by}{{\bf y}}

\newcommand{\ost}{_{\rm ost}}

\definecolor{nblue}{rgb}{0.06,0.3,0.73}
\definecolor{nblack}{rgb}{0,0,0}
\definecolor{nred}{rgb}{0.9,0.1,0.1}
\definecolor{nmagenta}{rgb}{0.7,0.0,0.3}
\newcommand{\blu}{\color{nblue}}

\newcommand{\blk}{\color{nblack}}

\definecolor{applegreen}{rgb}{0.55, 0.71, 0.0}
\newcommand{\grn}{\color{applegreen}}

\newcommand\stPW{\bgroup\markoverwith{\applegreen{\rule[0.5ex]{2pt}{0.4pt}}}\ULon}

\begin{document}

\title{A Retrodictive Approach to Quantum State Smoothing}

\author{Mingxuan Liu}
\affiliation{Centre for Quantum Technologies, National University of Singapore, 117543 Singapore, Singapore}
\author{Valerio Scarani}
\affiliation{Centre for Quantum Technologies, National University of Singapore, 117543 Singapore, Singapore}
\affiliation{Department of Physics, National University of Singapore, 2 Science Drive 3, Singapore 117542}
\author{Alexia Auff\`eves}
\affiliation{MajuLab, CNRS-UCA-SU-NUS-NTU International Joint Research Laboratory}
\affiliation{Centre for Quantum Technologies, National University of Singapore, 117543 Singapore, Singapore}
\date{\today}
\author{Kiarn T. Laverick}
\email{kiarn.l@nus.edu.sg}
\affiliation{MajuLab, CNRS-UCA-SU-NUS-NTU International Joint Research Laboratory}
\affiliation{Centre for Quantum Technologies, National University of Singapore, 117543 Singapore, Singapore}
\date{\today}

\begin{abstract}
Smoothing is a technique for estimating the state of an imperfectly monitored open system by combining both prior and posterior measurement information. In the quantum regime, current approaches to smoothing either give unphysical outcomes, due to the non-commutativity of the measurements at different times, or require assumptions about how the environment is measuring the system, which with current technology is unverifiable. We propose a novel definition of the smoothed quantum state based on quantum Bayesian retrodiction, which mirrors the classical retrodictive approach to smoothing. This approach always yields physical results and does not require any assumption on the environment. We show that this smoothed state has, on average, greater purity than the state reconstructed using just the prior information. Finally, we make a connection with the smoothing theory of Guevara and Wiseman in a well-studied regime, and describe from a purely quantum perspective how it conditions on the posterior information.



\end{abstract}
\maketitle

Smoothing \cite{Rauch63,FraPot69,Weinert01,Sarkka13}, and a related technique called filtering \cite{Kalman60,Kushner64,Jac96,Jaz07}, are estimation techniques that apply to dynamic open systems under continuous-in-time measurements, aiming at constructing an estimate of the system state at any time $t$ based the measurement record. Filtering takes into account only the record prior to $t$, while smoothing uses the record both prior and posterior to $t$. Thus smoothing generally yields a state with a lower entropy than filtering due to the additional measurement information. Computing the smoothed state in the classical setting is rather simple. By contrast, defining the analogue of the smoothed state in quantum systems has been the subject of debate over the past decade and a half \cite{Tsa-PRL09,Tsa-PRA09, CDJ13, Web14, GJM13, GamMol14, GueWis15, Budini17, CGLW21, LWCW23}.

The first attempt at defining a smoothed quantum state can be traced back to well before the modern debate \cite{Watanabe55,Watanabe56,ABL64}. The explicit connection to smoothing was made by Tsang in 2009 \cite{Tsa-PRA09}. This `state' (first made explicit in the Supplementary of Ref.~\cite{GJM13} and latter symmetrized in Ref.~\cite{LCW-PRA21}), however, yields unphysical results in general. The reason for this is the non-commutativity of the operators describing the future measurement outcomes and the quantum analogue of the filtered state. This operator has been referred to as the smoothed weak-value state, due to its connections to the anomalous weak values of Aharonov {\em et al.}~\cite{AAV88}.

The current approach to quantum state smoothing was formulated in 2015 by Guevara and Wiseman \cite{GueWis15}. They first noticed that the smoothing problem only makes sense when there is still information missing from the filtered state (\ie, it is not pure). They then defined the quantum state of maximal knowledge that one is aiming to estimate, called the {\em true} quantum state of the system. To do so, they consider an open quantum system undergoing imperfect continuous-in-time measurement by Alice, who obtains a record ${\bf O}$. A second observer, Bob, measures the remainder (or some subset thereof) of the environment, obtaining a record ${\bf U}$ that is unknown to Alice. The true state at time $t$ is defined as the filtered quantum state obtained by conditioning on the past records, $\past{\bf O}_t$ and $\past{\bf U}_t$. With the true state defined, the standard filtered quantum state from Alice's perspective (as given by quantum trajectory theory \cite{Bel99,Carmichael93,WisMil10}) is a classical mixture over the set of possible true states, specified by Alice's past record $\past{\bf O}_t$. Finally, they defined the smoothed quantum state similarly: a classical mixture over the set of possible true states, specified by Alice's past \textrm{and future} records $\both{\bf O}$. It is easy to see that this state is always physical. 

Numerous properties of the Guevara-Wiseman approach have since been discovered, from the optimality \cite{LGW-PRA21} of the state to the best measurement unravelings for smoothing \cite{CGW19,LCW-PRA21}. The approach has also been applied to the specific class of linear Gaussian quantum systems \cite{LCW19}. However, their smoothed quantum state depends on the particular measurement unraveling chosen by Bob \cite{GueWis15,CGW19,LCW-PRA21,LWCW23}. The approach thus applies exactly only in the rather artificial setting, in which there are actually two observers measuring the totality of the system. In the more usual setting where the role of Bob is played by the environment, the definition of the true state requires committing to a definition of the environment's true behaviour, which is an extra assumption. In either case, computing the smoothed state requires going through all possible trajectories, which can only be done{\blk, in general,} approximately and numerically.

In this Letter, we provide an alternative resolution to the debate. We derive a novel solution to the quantum smoothing problem, which is always a valid quantum state and does not require any Bob or any description of the {\blk remaining} environment. The approach we take is based on an alternative way of looking at the classical smoothing theory as a form of Bayesian retrodiction. For retrodiction we use the Petz recovery map \cite{petz1986,petz1988}. The resulting smoothed state has a very simple analytical, closed-form expression, first obtained by Fuchs as the analog of the Bayes' rule for a quantum measurement \cite{Fuchs03}; thus, we call our construction the \textit{Petz-Fuchs smoothed quantum state}. It also satisfies all the desirable properties of a quantum state smoothing theory \cite{LCW-QS21}. On the canonical example of a resonantly driven qubit coupled to a thermal bosonic environment \cite{WallsMilb94,WisMil10}, we show that the Petz-Fuchs smoothed state, on average, yields a purer estimate than filtering, indicating that it, in general, gives a more refined reconstruction of the dynamics than does the filtered state. We conclude by connecting our approach with previous ones, notably the Guevara-Wiseman approach. 


{\em Classical Retrodictive Smoothing.}---We begin by reviewing the classical theory of smoothing, highlighting its connection to Bayesian retrodiction. Consider an open classical dynamical system described by a vector of parameters $\bx$. For simplicity, we will assume that the elements $x_i$ are discrete variables taking values in a set ${\mathbb X}$. The value of $\bx$ is stochastic and generally unknown due to the interactions with the environment, therefore, we can only describe the system probabilistically $\wp(\bx;t)$. The dynamics of this {\em classical state} $\wp(\bx;t)$ are assumed to be Markovian, governed by the rate equation \cite{GardinerBook,Jaz07,SpeChu08}
\beq
\wp(\bx';t + \dd t) = \sum_{\bx \in {\mathbb X}} \varphi(\bx';t+\dd t|\bx; t)\wp(\bx;t)\,,
\eeq
with initial condition $\wp(\bx;t_0)$. The dynamical matrix must satisfy the completeness condition $\sum_{\bx'\in{\mathbb X}}\varphi(\bx';t+\dd t|\bx; t) = 1$. Without any additional information about the system, this is the most accurate description possible. 

If we have access to the outcomes of a continuous-in-time measurement, generally imperfect, we can refine our description of the system by conditioning the dynamics on the measurement outcomes $\by_t$ obtained at time $t$. The conditioning of the state at time $t$ on the measurement information is given via Bayes' theorem \cite{Jaz07,SpeChu08}, $\tilde{\wp}(\bx; t|\by_t) = \wp(\by_t|\bx;t) \wp(\bx;t)$. Throughout this paper, because of the linearity of the expressions, we will work with unnormalized states, denoted by $\tilde{\bullet}$. Though negligible in many cases, even in classical physics the measurement causes a back-action on the system dynamics. Thus, the dynamical update in general depends on the outcome $\by_t$, with the stochastic map of back-action $\varphi_{\by_t}(\bx';t+\dd t|\bx; t)$ also satisfying a completeness condition. Repeatedly conditioning the state based on the current measurement outcome and evolving leads to the general filtering equation
\beq\label{eq:cl_filt}
\tilde{\wp}\fil(\bx';t+\dd t) = \sum_{\bx\in{\mathbb X}} {\cal F}_{\by_t}(\bx';t+\dd t|\bx; t)\tilde{\wp}\fil(\bx;t)\,,
\eeq
with $\tilde{\wp}\fil(\bx;t_0) = \wp(\bx;t_0)$ and the (unnormalized) conditional forward map ${\cal F}_{\by_t}(\bx';t+\dd t|\bx; t) = \varphi_{\by_t}(\bx';t+\dd t|\bx; t)\wp(\by_t|\bx;t)$. It is easy to show that the norm of $\tilde{\wp}\fil(\bx;t):=\tilde{\wp}(\bx;t|\past{\bf O}_t)$ is $\wp(\past{\bf O}_t)$.

To obtain the smoothed state, we need to further condition the state on the future measurement record. The smoothed state, following Bayes' theorem, is given by ${\wp}\sm(\bx;t) := \wp(\bx;t|\both{\bf O}) \propto E\rfil(\bx;t)\tilde{\wp}\fil(\bx;t)$, where the {\em retrofiltered effect} (to use quantum terminology) is defined as $E\rfil(\bx;t) = \wp(\fut{\bf O}_t|\bx;t)${\blk, where $\fut{\bf O}_t$ denotes the future measurement record from $t$ onward}. While the Bayes' theorem approach is by far the simplest, we wish to take a deeper look into the dynamics than this form allows. It is easy to show \cite{SM} that the smoothed state evolves backwards-in-time from $\tilde{\wp}\fil(\bx;t)$ according to 
\beq\label{eq:cl_sm}
\tilde{\wp}\sm(\bx;t) = \sum_{\bx' \in {\mathbb X}} {\cal R}_{\by_t}^{\tilde{\wp}\fil(\bx;t)}(\bx;t|\bx'; t + \dd t) \tilde{\wp}\sm(\bx';t+\dd t)\,,
\eeq
where the ${\cal R}_{\by_t}^{\tilde{\wp}\fil(\bx;t)}(\bx;t|\bx'; t + \dd t)$ is the conditional retrodictive map. With this, we can see that smoothing is equivalent to Bayesian retrodiction \cite{Watanabe55, BusSca21} {\blk conditioned on the future measurement outcomes}. Importantly, we note that the conditional retrodictive map is constructed based not only on the forward map ${\cal F}_{\by_t}$ but also on a probability distribution ${\tilde{\wp}\fil(\bx;t)}$, referred to as a {\em reference prior}. This reference prior is only important when the forward map ${\cal F}_{\by_t}$ is irreversible{\blk, in the sense that some information is lost in the update}. In these instances, the retrodictive map ${\cal R}_{\by_t}^{\tilde{\wp}\fil(\bx;t)}$ will bias the evolution towards the reference prior ${\tilde{\wp}\fil(\bx;t)}$.

{\em Quantum Retrodictive Smoothing.}---In the quantum case, we are interested in an analogous system to the classical one, that of a Markovian open quantum system. By Markovian, we mean in the sense that the dynamics of the quantum system are described by a Lindblad master equation \cite{BrePet06,WisMil10}. However, for simplicity, we can consider the dynamics of the system via $\rho(t+\dd t) = \hat{\cal E}[\rho(t)]$, where $\hat{\cal E}$ is a completely-positive trace-preserving (CPTP) map (the quantum analogue of $\varphi$) and the initial condition is taken to be $\rho(t_0) = \rho_0$. Without any additional information, this is the best description of the quantum system.


Analogous to the classical case, we will consider an imperfect continuous-in-time weak measurement of the environment. By performing a Kraus decomposition, in terms of the measurement operators $\hat{M}_{\by_t}$, on the unconditional channel $\hat{\cal E}$, we can unravel the dynamics into a collection of conditional forward maps $\hat{\cal F}_{\by_t}$, \ie, $\hat{\cal E} = \sum_{\by_t}\hat{\cal F}_{\by_t}$, where $\by_t$ denotes a measurement outcome and $\hat{\cal F}_{{\bf y}_t}(\bullet) = \sum_{\ell=0}^L \hat{M}_{{\bf y}_t}\hat{K}_\ell\bullet \hat{K}_\ell\dg\hat{M}_{{\bf y}_t}\dg$. The operator $\hat{K}_\ell$ describes any remaining dissipation due to measurement inefficiency. Chaining these conditional forward maps together based on the realized measurement outcomes via
\beq\label{q_fil}
\tilde{\rho}\fil(t+\dd t)=\hat{\cal F}_{{\bf y}_t}[\tilde{\rho}\fil(t)]\,,
\eeq
yields a quantum trajectory \cite{Carmichael93}, also known as the filtered quantum state \cite{Bel99}. As in the classical case, we will work with the unnormalized state. 

The quantum retrofiltered effect $\hat{E}\rfil(t)$ \cite{GJM13} is a positive operator-valued measure element that describes the future measurement outcomes. Dynamically, this effect evolves backwards-in-time from a final uninformative effect $\hat{E}\rfil(T) \propto \mathds{1}$ via
\beq
\hat{E}\rfil(t)=\hat{\cal F}_{{\bf y}_t}^\dagger[\hat{E}\rfil(t+\dd t)].
\eeq
{\blk Note that the retrofiltered effect, when normalized, has the properties of a valid quantum state, \ie, Hermitian and positive semidefinite; even then, one should not think of this as a traditional quantum state.} With $\tilde{\rho}\fil(t)$ and $\hat{E}\rfil(t)$ defined, the classical theory suggests that the smoothed state should be their product. Here is where non-commutativity spoils the game: neither the product in either order, nor the symmetrized version, produce a physical quantum state in general. 

To follow the retrodictive approach of the classical theory we need a method to reverse the conditional evolution of the filtered state.
Quantum reverse maps that approximately recover the input state after an {\blk informationally} irreversible process are an active subject of research \cite{ManJor19,EloJor19,SS23,PF23}. The Petz recovery map \cite{petz1986,petz1988} has been shown to be a unique candidate \cite{CS23,PB23,CLS24} as the analogy of classical Bayesian retrodiction. It has been developed not only for logical inference, but also as a protocol to control noise \cite{BK02,NHK10,KMK22}. For a given CPTP map $\hat{\blk \cal E}$ and a reference prior state $\gamma$, the usual Petz recovery map is defined as $\hat{\cal R}^{\gamma}_{\hat{\blk \cal E}}[\bullet]={\gamma}^{1/2}\hat{\blk \cal E}^\dagger\left[\hat{\blk \cal E}[\gamma]^{-1/2}\bullet \hat{\blk \cal E}[\gamma]^{-1/2}\right]{\gamma}^{1/2}$. 

We work with the Petz maps of conditional CP maps $\hat{\blk \mathcal{E}}=\hat{\cal F}_{\by_t}$, and write them $\hat{\cal R}_{\by_{t}}^{\gamma}$ to avoid excessive subscripts. With these quantum conditional retrodictive channels in hand, analogously to the classical case, we define a smoothed quantum state dynamically, via
\beq\label{eq:q_retro_sm}
\tilde{\rho}\sm(t) = \hat{\cal R}_{\by_{t}}^{\tilde{\rho}\fil(t)}[\tilde{\rho}\sm(t+\dd t)]\,,
\eeq
with final condition $\tilde{\rho}\sm(T) = \tilde{\rho}\fil(T)$. This dynamical equation is simple to solve, making use of the composability of the Petz map \cite{CLS24}. That is, for a composite channel $\hat{\cal E} = \hat{\cal E}_2\circ\hat{\cal E}_1$, it holds $\hat{\cal R}_{\hat{\cal E}_1}^\gamma \circ \hat{\cal R}_{\hat{\cal E}_2}^{\sigma} = \hat{\cal R}_{\hat{\cal E}}^{\gamma}$ with $\sigma = \hat{\cal E}_1[\gamma]$. In our case, this property is satisfied since our reference priors satisfy $\tilde\rho\fil(t+\dd t) = \hat{\cal F}_{\by_t}[\tilde{\rho}\fil(t)]$. Thus the smoothed state at any time $t$ can be expressed as the application of a single conditional retrodictive channel, \ie, $\tilde\rho\sm(t) = \hat{\cal R}_{\fut{\bf O}_t}^{\tilde{\rho}\fil(t)}[\tilde{\rho}\fil(T)]$. Finally, noticing that $\tilde{\rho}\fil(T) = \hat{\cal F}_{\fut{\bf O}_t}[\tilde{\rho}\fil(t)]$, we obtain
\beq \label{eq:q_sm}
\tilde{\rho}\sm(t) = \sqrt{\tilde{\rho}\fil(t)}\hat{E}\rfil(t)\sqrt{\tilde{\rho}\fil(t)}\,.
\eeq
See \cite{SM} for a detailed derivation, as well as an alternative one using classical registers and CPTP maps. In this form, the smoothed state appears as an application of Fuchs' quantum Bayes rule \cite{Fuchs03}, justifying the name \textit{Petz-Fuchs smoothed state} that we chose.

It is straightforward that this state is always a valid quantum state conditioned on any past and future measurement information. To be a proper smoothed state, it should satisfy additional properties \cite{LCW-QS21}. First, it should revert to classical smoothing if the quantum system is effectively classical. Second, averaging over all possible future records $\fut{\bf O}_t$ should give back the filtered state. We show that both of these properties are satisfied in \cite{SM}. Importantly, notice that, since $\tilde{\rho}\fil(t)$ and ${\hat E}\rfil(t)$ are independent of how the remaining information in the environment is unraveled, this smoothed state is also independent of any commitment to a ``true'' unraveling. {\blk One consequence of this is that the Petz-Fuchs smoothed state does not require the simulation numerous possible ``true'' trajectories to be computed, making it more computationally efficient to compute over the Guevara-Wiseman smoothed state. }

{\em Example: the optical Bloch equations.}---To illustrate the Petz-Fuchs smoothing on a concrete example, we consider a classically driven qubit coupled to a thermal bosonic field. This scenario is ubiquitous in quantum optics and is described by the optical Bloch equations \cite{WallsMilb94, WisMil10}
\beq\label{eq:ME}
\frac{\dd\rho}{\dd t} = -i[\hat{H}, \rho] + \gamma (\bar{n} + 1){\cal D}[\s{-}]\rho + \gamma\bar{n}{\cal D}[\s{+}]\rho\,.
\eeq
Here $\s{x,y,z}$ are the usual Pauli operators, $\s{+,-}$ are the Pauli raising and lowering operators, respectively, $\hat{H} = \Omega\s{y}/2$ is the Hamiltonian that results from a classical driving of the qubit with Rabi frequency $\Omega$, $\gamma$ is the dissipation rate and $\bar{n}$ the average photon number in the thermal environment. The dissipation superoperator is ${\cal D}[\hat{c}]\bullet = \hat{c}\bullet\hat{c}\dg - \half\{\hat{c}\dg\hat{c},\bullet\}$. 

For the measurement, we {\blk consider monitoring only the photon emitted from the system, described by the dissipator ${\cal D}[\s{-}]$. Specifically,} we consider three possible ways to monitor the {\blk emitted} photons: 1) photon detection (jumps), 2) X-homodyne detection (diffusive) and 3) Y-homodyne detection (diffusive). Since there is a second dissipation channel {\blk ${\cal D}[\s{+}]$ (describing incoherent photon absorption)} that is unmeasured, the system will always be imperfectly monitored irrespective of the proportion of light detected from the emission channel. {\blk This means that the resulting filtered state will always be missing some information, i.e., having a non-unit purity, which smoothing can recover}. For simplicity, we assume in each case that the monitoring of the emitted light is perfect. 

\begin{figure}[h]
\includegraphics[scale=0.45]{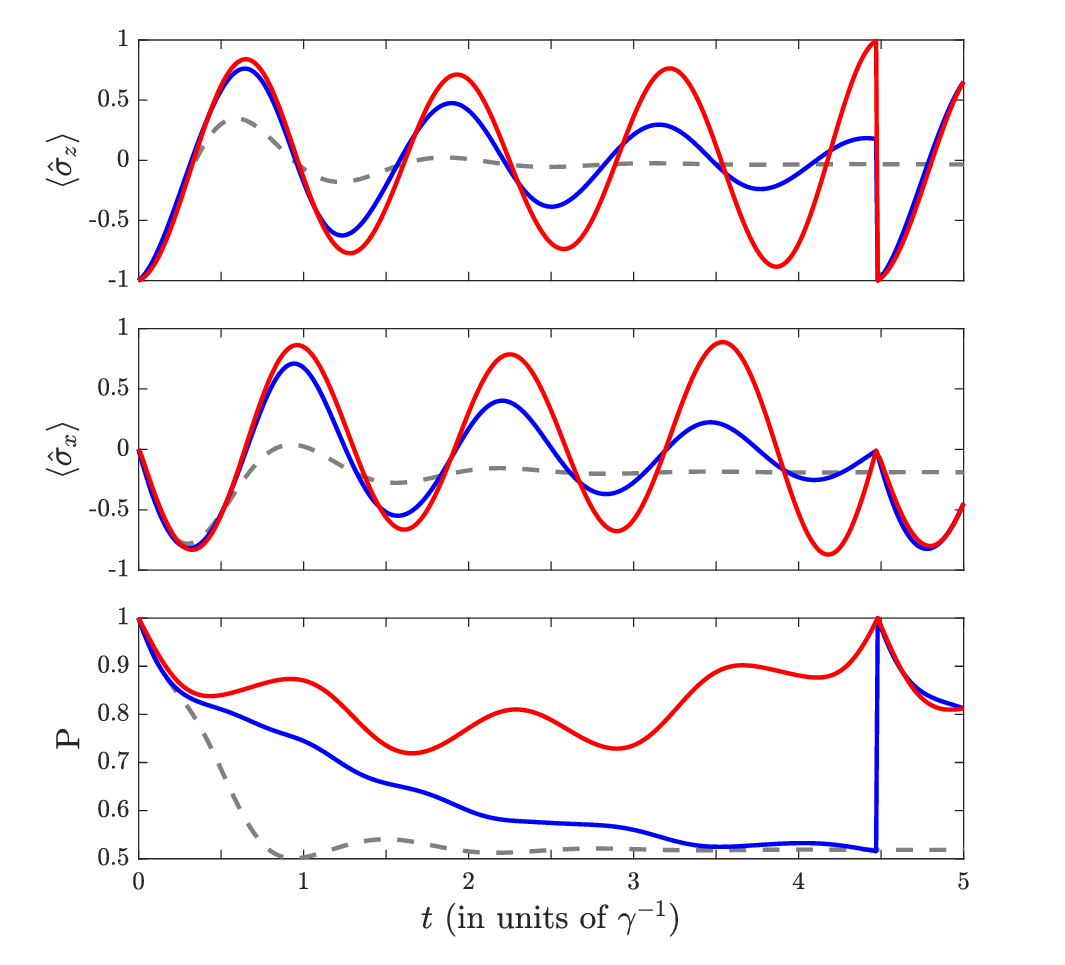}
\caption{A sample trajectory of the filtered state (blue) and the smoothed state (red) in terms of their Bloch components as well as the corresponding purity of the state. For reference, the dashed gray line shows the trajectory of unconditional state that evolves according to Lindblad master equation. Note that in this case $\ex{\s{y}}(t) = 0$ for all three states. For this trajectory, only a single photon is detected during the evolution, at $t\approx4.5\gamma\inv$. Here, $\Omega=5\gamma$ and $\bar{n} = 0.5$.}
\label{fig:Trajectories}
\end{figure}

In Fig.~\ref{fig:Trajectories}, we show a single realization, in the case of photon detection, of the Petz-Fuchs smoothed state, alongside the filtered state and the unconditional state (the solution to \erf{eq:ME}). Unsurprisingly, the smoothed state is purer than the filtered state over the initial times, due to the knowledge that the qubit has not undergone spontaneous emission until almost the final time. Consequently, in the smoothed state, the Rabi dynamics of the qubit has almost been fully revived; there is still a dip in purity at intermediate times, where the dissipation due to incoherent absorption dominates because the jump is too far in the future. After detection, there are times at which the smoothed state is less pure.

\begin{figure}[t!]
\includegraphics[scale = 0.36]{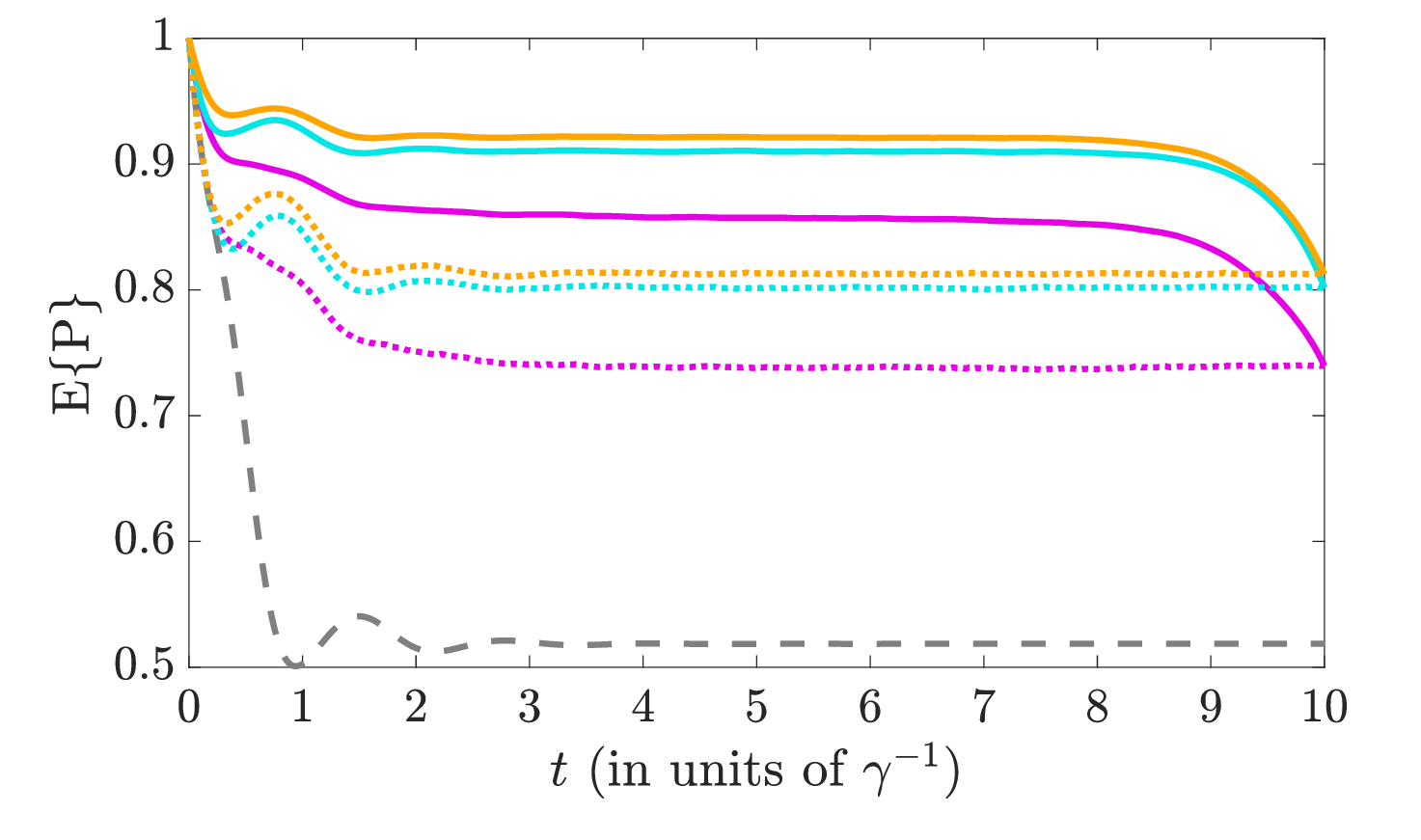}
\caption{The average purity of smoothed state (solid lines) and corresponding filtered state (dotted lines) for photon counting detection (magenta), X-Homodyne (cyan) and Y-Homodyne (orange) detection respectively. The purity of the unconditioned state is plotted (dashed grey line) for reference. We see that, in the steady state region $t\in[4\gamma\inv, 7.5\gamma\inv]$, that on average the smoothed state outperforms the filtered state in terms of purity for all measurements considered. The average was calculated with 30000 randomly generated past-future records. Here, $\Omega = 5\gamma$ and $\bar{n} = 0.5$. }
\label{fig:avg_purity}
\end{figure}

To show that the smoothed state offers an actual improvement over the filtered state, irrespective of the particular record obtained or boundary effects, we consider the average purity over all trajectories, \ie, ${\rm E}\{{\rm P}\} = \sum_{\both{\bf O}} \wp(\both{\bf O}) \Tr[\rho^2]$. In Fig.~\ref{fig:avg_purity}, we see that in the steady state regime, $t\in[4\gamma\inv,7.5\gamma\inv]$, the smoothed state (solid lines) is always purer than the filtered state (dashed lines) for all measurements considered. Note that for a realistic measurement, \ie, detecting some non-unit proportion of the radiated light, the purity improvement given by the smoothed state will be smaller. 

Interestingly, the greatest relative improvement is given by photon counting detection, while the two homodyne detections show approximately the same improvement. Moreover, it is surprising that Y-homodyne gives a better absolute purity than X-homodyne, since in the latter case the qubit is confined to the $x$-$z$ plane circle, and gaining information about the $x$-component would seemingly provide the greatest information about the state. {\blk However, what is likely happening is that, since the Y-homodyne measurement is orthogonal to the plane in which the dynamics are occurring, the measurement backaction induced by this measurement has minimal disturbance on the dynamics, leading to a slightly larger information gain, on average, than X-homodyne measurements, where the backaction occurs in the $x$-$z$ plane.}

{\em Comparison to Guevara-Wiseman smoothing.}---
One may wonder whether the Guevara-Wiseman smoothed state can also be understood in terms of a retrodictive approach to smoothing. As it turns out, the answer to this question is yes, although only in a particular, often considered, regime. 

In the Guevara-Wiseman scenario there are two observers Alice and Bob, each measuring some unique subset of the environment. Given the past measurement records $\past{\bf O}_t$ and $\past{\bf U}_t$, respectively, the true state is defined as the quantum state conditioned on both: $\rho\god(t) = \rho_{\past{\bf O}_t,\past{\bf U}_t}(t)$. We shall also use the unnormalized $\tilde\rho\god(t) = \wp(\past{\bf O}_t,\past{\bf U}_t)\rho\god(t)$ which embeds the likelihood of each true state occurring.

Since Alice does not have access to Bob's record $\past{\bf U}_t$, the best she can do is to describe the state based solely on her measurement record ${\bf O}$. For her filtering, since the true state is already conditioned on her past measurement record, all that is required is to average over the possible ${\past{\bf U}_t}$, giving $\tilde{\rho}\fil(t) = \sum_{\past{\bf U}_t}\tilde{\rho}\god(t)$. As Bob's measurement amounts to a particular choice for the Kraus operators $\{\hat{K}_\ell\}$, this definition returns the standard filtered state \erf{q_fil}. As for her smoothing, an intuitive application of our approach is: first, we use retrodiction to condition true state on Alice's future record $\fut{\bf O}_t$ via \erf{eq:q_sm}; then we average over the past unobserved record. This yields
\beq\label{eq:Petz-Wiseman_sm}
\tilde{\rho}\sm(t) = \sum_{\past{\bf U}_t}\wp(\past{\bf U}_t|\past{\bf O}_t)\sqrt{\rho\god(t)}\hat{E}\rfil(t)\sqrt{\rho\god(t)}
\eeq
(notice that the true state in this equation is normalized). The direct application of the Petz map gives the same result \cite{SM}, vindicating our intuition.


At the moment, we have applied our smoothing approach to the Guevara-Wiseman scenario, but have not yet established a connection to the Guevara-Wiseman smoothed state, defined as ${\rho}\sm^{\rm GW}(t) = \sum_{\past{\bf U}_t} \wp(\past{\bf U}_t|\both{\bf O})\rho\god(t)$ \cite{GueWis15}. We are going to prove that the two definitions yield identical smoothed states when the true state is pure $\rho\god(t) = \op{\psi\god(t)}{\psi\god(t)}$, as is often taken to be  \cite{GueWis15,CGLW21,LWCW23}. Indeed, noting that $\bra{\psi\god(t)}\hat{E}\rfil(t)\ket{\psi\god(t)} = \wp(\fut{\bf O}_t|\past{\bf O}_t,\past{\bf U}_t)$, we see that \erf{eq:Petz-Wiseman_sm}, after normalization, reduces to $\rho\sm(t) = \sum_{\past{\bf U}_t} \wp(\past{\bf U}_t|\both{\bf O})\op{\psi\god(t)}{\psi\god(t)}$, \ie, the Guevara-Wiseman smoothed state. Interestingly, this form of the Guevara-Wiseman smoothed state provides the first insight (from a purely quantum mechanical perspective) as to how the future information is taken into account, whereas previously this was hidden within the classical probability. Moreover, we can see the reason that the smoothed state depends on Bob's choice in measurement is due to the non-linearity in the true state (this holds more generally to since $\wp(\past{\bf U}_t|\both{\bf O}) \propto \Tr[\hat{E}\rfil(t)\rho\god(t)]\wp(\past{\bf U}_t|\past{\bf O})$). Further conclusions drawn about Guevara-Wiseman smoothing from this form should, however, be taken with a grain of salt as, when $\rho\god(t)$ is mixed, \erf{eq:Petz-Wiseman_sm} is a completely different way of conditioning on the future information. 


{\em Conclusion.}---To conclude, we have developed a novel quantum state smoothing theory centered around the Petz recovery map. Importantly, unlike previous solutions to this problem, our approach always yields a physical quantum state as its solution as well as not requiring any additional assumptions on how the environment effectively measures the system, greatly widening the potential applicability of the smoothing technique in open quantum systems. Through an example, we have shown that this Petz-Fuchs smoothed state yields an improvement over filtering, on average, in terms of purity recovery. Finally, we have shown that the Guevara-Wiseman approach to smoothing, in the case where the underlying true state of the system is pure, is equivalent to our Petz recovery map, allowing us to understand, on a purely quantum level, how the Guevara-Wiseman smoothed state conditions the state on the future measurement record. 

A question for future work is to investigate whether the Petz-Fuchs smoothed state is optimal in some sense, as all the other cases (classical smoothing, Guevara-Wiseman smoothing and even weak-valued smoothing) are optimal Bayesian estimators \cite{CGLW21} for some cost function. Additionally, it would be interesting to apply this smoothing technique to the regime of linear Gaussian quantum systems, as they provide a natural test bed for theories due to their simplicity while still describing a wide range of physical systems \cite{Bra05,BowMil16}. In particular, it is known that for classical linear Gaussian systems, the retrodictive approach to smoothing leads to the Rauch-Tung-Striebel smoothing equations \cite{Rauch63}, whereas in the quantum case it cannot reduce to these equations as they are known \cite{Tsa-PRA09, Laverick21} to give unphysical results. Thus, in the linear Gaussian regime, we are likely to learn some fundamental differences between classical and quantum retrodiction. 

{\em Acknowledgments} --- We would like to thank Ge Bai and Howard M. Wiseman for many helpful discussions. This work is supported by the National Research Foundation, Singapore and A*STAR under its CQT Bridging Grant. V.S.~acknowledges support from the Ministry of Education, Singapore, under the Tier 2 grant ``Bayesian approach to irreversibility'' (Grant No.~MOE-T2EP50123-0002).

%

\newpage
\begin{widetext}
\section{Supplementary Material: A Retrodictive Approach to \\Quantum State Smoothing}
\renewcommand{\theequation}{S.\arabic{equation}}
\setcounter{equation}{0}
\subsection{Derivation of \erf{eq:cl_sm} from Bayes' theorem}
In order to show that the smoothed state obtained via Bayes' theorem is identical to \erf{eq:cl_sm}, we first need to derive the evolution equation for the 
retrofiltered effect. To do this, let us consider the unnormalized filtered state $\tilde{\wp}\fil(\bx;t) = \wp(\bx,\past{\bf O}_t; t)$. When multiplying this 
distribution by $E\rfil(\bx;t):=\wp(\fut{\bf O}_t|\bx;t)\equiv \wp(\fut{\bf O}_t|\bx,\past{\bf O}_t;t)$, where the final equivalence is due to the Markovianity of 
the system, the resulting unnormalized smoothed state is 
\beq
\tilde{\wp}\sm(\bx;t) = E\rfil(\bx;t)\tilde{\wp}\fil(\bx;t) = \wp(\bx,\both{\bf O};t)\,.
\eeq
Summing over $\bx \in {\mathbb X}$, we obtain
\beq
\wp(\both{\bf O}) = \sum_{\bx\in{\mathbb X}} E\rfil(\bx;t)\tilde{\wp}\fil(\bx;t)\,.
\eeq
Since this distribution is time-independent, it must be that
\beq
\begin{split}
\sum_{\bx\in{\mathbb X}}E\rfil(\bx;t)\tilde{\wp}\fil(\bx;t) &= \sum_{\bx'\in{\mathbb X}} E\rfil(\bx';t+\dd t)\tilde{\wp}\fil(\bx';t + \dd t)\\
& = \sum_{\bx'\in{\mathbb X}} E\rfil(\bx';t+\dd t) \sum_{\bx\in{\mathbb X}}{\cal F}_{\by_t}(\bx';t+\dd t|\bx;t) \tilde{\wp}\fil(\bx;t)\\
& = \sum_{\bx\in{\mathbb X}} \left\{ \sum_{\bx'\in{\mathbb X}}{\cal F}_{\by_t}(\bx';t+\dd t|\bx;t) E\rfil(\bx';t+\dd t)\right\} \tilde{\wp}\fil(\bx;t)\,,
\end{split}
\eeq
where, in the second line, we have used \erf{eq:cl_filt}. From this, we can see that the retrofiltered effect evolves backwards-in-time according to 
\beq\label{eq:retr_evo}
E\rfil(\bx;t) = \sum_{\bx'\in{\mathbb X}}{\cal F}_{\by_t}(\bx';t+\dd t|\bx;t) E\rfil(\bx';t+\dd t)\,,
\eeq
where the final condition is taken to be $E\rfil(\bx;T) \propto 1$, as at the final time $\past{\bf O}_T = \both{\bf O}$.

With the dynamical equation for the retrofiltered effect acquired, we can now derive \erf{eq:cl_sm}. Let us consider
\beq
\tilde{\wp}\sm(\bx;t) = E\rfil(\bx;t) \tilde{\wp}\fil(\bx;t)\,.
\eeq
Substituting in \erf{eq:retr_evo}, we obtain
\beq
\tilde{\wp}\sm(\bx;t) =  \sum_{\bx'\in{\mathbb X}}{\cal F}_{\by_t}(\bx';t+\dd t|\bx;t) E\rfil(\bx';t+\dd t) \tilde{\wp}\fil(\bx;t)\,.
\eeq
Realizing that $E\rfil(\bx;t+\dd t) = \tilde{\wp}\sm(\bx;t+\dd t)/\tilde{\wp}\fil(\bx;t+\dd t)$, we have
\beq
\begin{split}
\tilde{\wp}\sm(\bx;t) &=  \sum_{\bx'\in{\mathbb X}}{\cal F}_{\by_t}(\bx';t+\dd t|\bx;t) \frac{\tilde{\wp}\sm(\bx';t+\dd t)}{\tilde{\wp}\fil(\bx';t+\dd t)} \tilde{\wp}\fil(\bx;t)\\
& = \sum_{\bx'\in{\mathbb X}} \frac{{\cal F}_{\by_t}(\bx';t+\dd t|\bx;t)\tilde{\wp}\fil(\bx;t)}{\sum_{\bx''\in{\mathbb X}}{\cal F}_{\by_t}(\bx';t+\dd t|\bx'';t)\tilde{\wp}\fil(\bx'';t)} \tilde{\wp}\sm(\bx;t+\dd t)\,.
\end{split}
\eeq
By defining the reverse map as 
\beq\label{eq:cl_re_map}
{\cal R}_{\by_t}^{\tilde{\wp}\fil(\bx;t)}(\bx;t|\bx';t+\dd t) = \frac{{\cal F}_{\by_t}(\bx';t+\dd t|\bx;t)\tilde{\wp}\fil(\bx;t)}{\sum_{\bx''\in{\mathbb X}}{\cal F}_{\by_t}(\bx';t+\dd t|\bx'';t)\tilde{\wp}\fil(\bx'';t)}\,,
\eeq
we arrive at \erf{eq:cl_sm}.

\subsection{Detailed derivation of the Petz-Fuchs smoothed state \erf{eq:q_sm}}
\label{App:derive_PF}
In this section we will derive \erf{eq:q_sm} from \erf{eq:q_retro_sm}. Let us begin with the explicit form of the conditional retrodictive quantum channel,
\beq\label{eq:cond_rev_map}
\hat{\cal R}^{{\tilde{\rho}\fil(t)}}_{{\bf y}_t}[\bullet]=\sqrt{\tilde{\rho}\fil(t)}\hat{\cal F}_{{\bf y}_t}^\dagger\left[\frac{1}{\sqrt{\hat{\cal F}_{{\bf y}_t}[{\tilde{\rho}\fil(t)}]}}\bullet \frac{1}{\sqrt{\hat{\cal F}_{{\bf y}_t}[{\tilde{\rho}\fil(t)}]}}\right]\sqrt{\tilde{\rho}\fil(t)}\,,
\eeq
where $\hat{\cal F}_{{\bf y}_t}(\bullet) = \sum_{\ell=1}^L \hat{M}_{{\bf y}_t}\hat{K}_\ell\bullet \hat{K}_\ell^\dagger \hat{M}_{{\bf y}_t}^\dagger$. 
The Petz-Fuchs smoothed quantum state is defined via
\beq
    \tilde{\rho}\sm(t)=\hat{\cal R}_{\by_t}^{\tilde{\rho}\fil(t)}[\tilde{\rho}\sm(t+\dd t)],
\eeq
where $\tilde{\rho}\sm(T) = \tilde{\rho}\fil(t)$. As this is an iterative equation, we can also express the smoothed state at time $t$ as 
\beq
\tilde{\rho}\sm(t) =\hat{\cal R}_{\by_t}^{\tilde{\rho}\fil(t)}\left[\hat{\cal R}_{\by_{t+\dd t}}^{\tilde{\rho}\fil(t+\dd t)}[\tilde{\rho}\sm(t+2\dd t)]\right]\,.
\eeq
Using \erf{eq:cond_rev_map}, we have
\beq
\begin{split}
\tilde{\rho}\sm(t) &= \sqrt{{\tilde{\rho}\fil(t)}}\hat{\cal F}_{{\bf y}_t}^\dagger\left[\frac{1}{\sqrt{\hat{\cal F}_{{\bf y}_{t}}[{\tilde{\rho}\fil(t)}]}}\hat{\cal R}_{\by_{t+\dd t}}^{\tilde{\rho}\fil(t+\dd t)}[\tilde{\rho}\sm(t+2\dd t)] \frac{1}{\sqrt{\hat{\cal F}_{{\bf y}_{t}}[{\tilde{\rho}\fil(t)}]}}\right]\sqrt{{\tilde{\rho}\fil(t)}}\\
&= \sqrt{{\tilde{\rho}\fil(t)}}\hat{\cal F}_{{\bf y}_t}^\dagger\left[\hat{\cal F}_{{\bf y}_{t+\dd t}}^\dagger\left[\frac{1}{\sqrt{\hat{\cal F}_{{\bf y}_{t+\dd t}}[{\tilde{\rho}\fil(t+\dd t)}]}}\tilde{\rho}\sm(t+2\dd t) \frac{1}{\sqrt{\hat{\cal F}_{{\bf y}_{t+\dd t}}[{\tilde{\rho}\fil(t+\dd t)}]}}\right]\right]\sqrt{{\tilde{\rho}\fil(t)}}\,,
\end{split}
\eeq
where, in the second line, we have simplified the expression by noting that $\tilde{\rho}\fil(t+\dd t) = \hat{\cal F}_{{\bf y}_t}[\tilde{\rho}\fil(t)]$.
Denoting $\hat{\cal F}_{{\bf y}_{t+\dd t}}\circ \hat{\cal F}_{{\bf y}_{t}}$ by $\hat{\cal F}_{{\bf y}_{t},{\bf y}_{t+\dd t}}$, we can further simplify this expression to 
\beq
\begin{split}
\tilde{\rho}\sm(t)&=\sqrt{{\tilde{\rho}\fil(t)}}\hat{\cal F}_{{\bf y}_{t},{\bf y}_{t+\dd t}}^\dagger\left[\frac{1}{\sqrt{\hat{\cal F}_{{\bf y}_{t},{\bf y}_{t+\dd t}}[{\tilde{\rho}\fil(t)}]}}\tilde{\rho}\sm(t+2\dd t) \frac{1}{\sqrt{\hat{\cal F}_{{\bf y}_{t},{\bf y}_{t+\dd t}}[{\tilde{\rho}\fil(t)}]}}\right]\sqrt{{\tilde{\rho}\fil(t)}}\\
&=\hat{\cal R}_{\by_t,\by_{t+\dd t}}^{\tilde{\rho}\fil(t)}[\tilde{\rho}\sm(t+2\dd t)]\,,
\end{split}
\eeq
where in the final line we have recognized that this map is just the conditional retrodictive quantum channel for the composite channel $\hat{\cal F}_{{\bf y}_{t},{\bf y}_{t+\dd t}}$.

While we have only illustrated the composability of the conditional retrodictive quantum channel for two sequential channels, it can easily be extended to the entire interval $[t,T)$. If we denote the composition of all conditional forward quantum channels in the interval $[t,T)$ as $\hat{\cal F}_{\fut{\bf O}_t}:=\hat{\cal F}_{{\bf y}_{T-\dd t}}\circ\hat{\cal F}_{{\bf y}_{T-2\dd t}}\circ\cdots\circ \hat{\cal F}_{{\bf y}_{t}}$, by the composability property, we end up with
\beq\label{eq:composed_petz}
\begin{split}
\tilde{\rho}\sm(t)&=\hat{\cal R}_{\fut{\bf O}_t}^{\tilde{\rho}\fil(t)}[\tilde{\rho}\sm(T)]\\
&=\sqrt{{\tilde{\rho}\fil(t)}}\hat{\cal F}_{\fut{\bf O}_t}^\dagger\left[\frac{1}{\sqrt{\hat{\cal F}_{\fut{\bf O}_t}[{\tilde{\rho}\fil(t)}]}}\tilde{\rho}\sm(T) \frac{1}{\sqrt{\hat{\cal F}_{\fut{\bf O}_t}[{\tilde{\rho}\fil(t)}]}}\right]\sqrt{{\tilde{\rho}\fil(t)}}\,.
\end{split}
\eeq
Lastly, remembering that $\tilde{\rho}\sm(T) = \tilde{\rho}\fil(T) = \hat{\cal F}_{\fut{\bf O}_t}[{\tilde{\rho}\fil(t)}]$, we obtain 
\beq
\begin{split}
\tilde{\rho}\sm(t) &= \sqrt{\tilde{\rho}\fil(t)}\hat{\cal F}_{\fut{\bf O}_t}^\dagger\left[\mathds{1}\right]\sqrt{\tilde{\rho}\fil(t)}\\
& = \sqrt{\tilde{\rho}\fil(t)}\hat{E}\rfil(t)\sqrt{\tilde{\rho}\fil(t)}.
\end{split}
\eeq
where, in the final line, we have recognized that $\hat{\cal F}_{\fut{\bf O}_t}^\dagger\left[\mathds{1}\right]=\hat{E}\rfil(t)$.

\subsection{Alternative derivation of the smoothed state with classical registers}

In what follows, we provide a method of applying the Petz recovery map through CPTP maps (as opposed to just CP maps) to derive the same result. We do this by introducing a classical register for the measurement outcomes into the problem. Let us begin with the initial quantum state $\rho_0$. After performing measurements continuously, with outcomes $\past{\bf O}_t$ recorded in classical registers, we obtain, at time $t$, the hybrid classical-quantum state
\beq
\tilde{\varrho}\fil(t)=\hat{\cal F}_{\past{\bf O}_t}[\rho_0]\otimes\op{\past{\bf O}_t}{\past{\bf O}_t}\,,
\eeq 
where $\hat{\cal F}_{\past{\bf O}_t}$ is the forward quantum operation conditioned on all the measurement outcomes in the interval $[t_0,t)$. Note, we are still working with unnormalized states here, purely for simplicity.

We can describe the evolution of the continuous-in-time measurement process from time $t$ to $T$, for any possible observed measurement record $\fut{\bf O}_t$, by the CPTP map ${\cal E}_{t\to T}[\bullet]=\sum_{\fut{\bf O}_t}(\hat{\cal F}_{\fut{\bf O}_t}\otimes {\cal I}_{C_{\past{\bf O}_t}})[\bullet]\otimes\op{\fut{\bf O}_t}{\fut{\bf O}_t}$, where ${\cal I}_{C_{\past{\bf O}_t}}$ is the identity channel on classical registers up to time $t$. Note, for maps that involve classical registers, we have omitted the $\hat{\bullet}$ accent to avoid confusion. This is the CPTP map we will use in the Petz recovery map to construct our retrodiction channel back to time $t$, \ie, ${\cal R}^{\tilde{\varrho}\fil(t)}_{{ \cal E}_{t\to T}}[\bullet]$. 

Now that we have the preliminaries out of the way, we can now apply our retrodiction channel to the hybrid filtered state at the final time, $\tilde{\varrho}\fil(T)=\hat{\cal F}_{\both{\bf O}}[\rho_0]\otimes\op{\both{\bf O}}{\both{\bf O}}$, which contains all the information about which measurement record was realized, to obtain our smoothed state. That is 
\beq\label{seq:CPTP_Petz}
    \tilde{\varrho}\sm(t)={\cal R}^{\tilde{\varrho}\fil(t)}_{{ \cal E}_{t\to T}}[\tilde{\varrho}\fil(T)]=\sqrt{\tilde{\varrho}\fil(t)}{\cal E}^\dagger\left[\frac{1}{\sqrt{{\cal E}[\tilde{\varrho}\fil(t)]}}\tilde{\varrho}\fil (T)\frac{1}{\sqrt{{\cal E}[\tilde{\varrho}\fil(t)]}}\right]\sqrt{\tilde{\varrho}\fil(t)},
\eeq
Using the orthogonality of the classical registers, we have
\beq
\frac{1}{\sqrt{{\cal E}[\tilde{\varrho}\fil(t)]}}=\sum_{\fut{\bf O}_t}\frac{1}{\sqrt{\hat{\cal F}_{\both{\bf O}}[\rho_0]}}\otimes \op{\both{\bf O}}{\both{\bf O}}\,,
\eeq
from which, we obtain
\beq
\begin{split}
    {\cal E}^\dagger\left[\frac{1}{\sqrt{{\cal E}[\tilde{\varrho}\fil(t)]}}\tilde{\varrho}\fil (T)\frac{1}{\sqrt{{\cal E}[\tilde{\varrho}\fil(t)]}}\right]&={\cal E}^\dagger[\mathds{1}_Q\otimes\op{\both{\bf O}}{\both{\bf O}}]\\&=\hat{\cal F}^\dagger_{\fut{\bf O}_t}[\mathds{1}_Q]\otimes\op{\past{\bf O}_t}{\past{\bf O}_t}\\
    &=\hat{E}\rfil(t)\otimes\op{\past{\bf O}_t}{\past{\bf O}_t},
\end{split}
\eeq
where in the final line we have recognized that $\hat{E}\rfil(t) = \hat{\cal F}^\dagger_{\fut{\bf O}_t}[\mathds{1}_Q]$, with $\mathds{1}_Q$ denoting the identity on for the quantum system.
Substituting this into \erf{seq:CPTP_Petz} yields 
\beq
\begin{split}
    \tilde{\varrho}\sm(t)=&\sqrt{\tilde{\varrho}\fil(t)}\left(\hat{E}\rfil(t)\otimes\op{\past{\bf O}_t}{\past{\bf O}_t}\right)\sqrt{\tilde{\varrho}\fil(t)}\\=&\sqrt{\hat{\cal F}_{\past{\bf O}_t}[\rho_0]}\hat{E}\rfil(t)\sqrt{\hat{\cal F}_{\past{\bf O}_t}[\rho_0]}\otimes\op{\past{\bf O}_t}{\past{\bf O}_t}.
\end{split}
\eeq
Finally, tracing out the classical register and using the fact that $\hat{\cal F}_{\past{\bf O}_t}[\rho_0]=\tilde{\rho}\fil(t)$, we obtain
\beq
\tilde{\rho}\sm(t)=\Tr_C[\tilde{\varrho}\sm(t)]=\sqrt{\tilde{\rho}\fil(t)}\hat{E}\rfil(t)\sqrt{\tilde{\rho}\fil(t)}.
\eeq

As an aside, instead of considering the smoothing setting, one can consider a simpler scenario involving a single measurement, where we are estimating the pre-measurement state given the measurement outcome. This problem has been studied for decades and a couple of proposals, all inspired by Bayes' theorem, were put forward \cite{Barnett2000, Fuchs03} and summarized in a rigorous math framework in Ref.~\cite{LS13}. It is easy to see that our result coincides with theirs in the single measurement scenario.

\subsection{Proof that \erf{eq:q_sm} satisfies the smoothing criteria}
In this section, we will prove that our retrodictive smoothed quantum state satisfies all the properties for a smoothed quantum state defined in Ref.~\cite{LCW-QS21}.\\

\underline{\bf Criteria 1:} {\em \blk The theory should give a single smoothed quantum state $\rho\sm$ 
analogous to the classical state $\wp\sm$, and not a pair of states, for example.}\\

This is evident from the definition in \erf{eq:q_sm}.\\

\underline{\bf Criteria 2:} {\em \blk The smoothed state $\rho\sm \equiv \rho_{\both{\bf O}}$ should 
reduce to its corresponding filtered state after averaging over all possible future (observed) measurement 
records given a past measurement record.} \\

The conditional average of the retrodicted smoothed state is
\beq
\sum_{\fut{\bf O}_t} \wp(\fut{\bf O}_t|\past{\bf O}_t)\rho\sm(t) = \sum_{\fut{\bf O}_t} \wp(\fut{\bf O}_t|\past{\bf O}_t) \frac{\sqrt{\rho\fil(t)}\hat{E}\rfil(t)\sqrt{\rho\fil(t)}}{\Tr[\hat{E}\rfil(t)\rho\fil(t)]}\,.
\eeq
Noting that, by definition, $\Tr[\hat{E}\rfil(t)\rho\fil(t)] = \wp(\fut{\bf O}_t|\past{\bf O}_t)$, we are left with 
\beq
\sum_{\fut{\bf O}_t} \wp(\fut{\bf O}_t|\past{\bf O}_t)\rho\sm(t) =\sum_{\fut{\bf O}_t} \sqrt{\rho\fil(t)}\hat{E}\rfil(t)\sqrt{\rho\fil(t)}\,.
\eeq
Using the fact that $\rho\fil(t)$ is independent of $\fut{\bf O}_t$ and that the retrofiltered effect must satisfy the completeness relation $\sum_{\fut{\bf O}_t} \hat{E}\rfil(t) = \sum_{\fut{\bf O}_t}\hat{\cal F}_{\fut{\bf O}_t}^\dagger(\mathds{1})={\mathds{1}}$, we have 
\beq
\sum_{\fut{\bf O}_t} \wp(\fut{\bf O}_t|\past{\bf O}_t)\rho\sm(t) = \rho\fil(t)\,.
\eeq

\underline{\bf Criteria 3:} {\em \blk The smoothed quantum state should reduce to its classical 
counterpart when the initial conditions, final conditions, and dynamics of the system can all be described 
probabilistically in a fixed basis.}\\

Under the constraints of the criterion, at any time, the filtered state and the retrofiltered effect can be written as $\rho\fil(t) = \sum_{x\in\mathbb X}\wp\fil(x;t)\ket{\psi_x}\bra{\psi_{x}}$ and $\hat{E}\rfil(t) = \sum_{x\in\mathbb X}E\rfil(x;t)\ket{\psi_x}\bra{\psi_{x}}$, where $\{\ket{\psi_x}\}$ is a fixed orthonormal basis. The retrodictive smoothed quantum state is thus
\begin{align}
\rho\sm(t) &= \frac{\sqrt{\rho\fil(t)}\hat{E}\rfil(t)\sqrt{\rho\fil(t)}}{\Tr[\sqrt{\rho\fil(t)}\hat{E}\rfil(t)\sqrt{\rho\fil(t)}]}\\
&= \frac{1}{\wp(\fut{\bf O}_t|\past{\bf O}_t)}\sqrt{\sum_{x\in{\mathbb X}}\wp\fil(x;t)\ket{\psi_x}\bra{\psi_x}}\sum_{x'\in{\mathbb X}}E\rfil(x';t)\ket{\psi_{x'}}\bra{\psi_{x'}}\sqrt{\sum_{x''\in{\mathbb X}}\wp\fil(x'';t)\ket{\psi_{x''}}\bra{\psi_{x''}}}\\
&= \frac{1}{\wp(\fut{\bf O}_t|\past{\bf O}_t)}\sum_{x,x',x''\in{\mathbb X}}\sqrt{\wp\fil(x;t)}E\rfil(x';t)\sqrt{\wp\fil(x'';t)}\ket{\psi_x}\ip{\psi_x}{\psi_{x'}}\ip{\psi_{x'}}{\psi_{x''}}\bra{\psi_{x''}}\\
&= \frac{1}{\wp(\fut{\bf O}_t|\past{\bf O}_t)} \sum_{x,x',x''\in{\mathbb X}}\sqrt{\wp\fil(x;t)}E\rfil(x';t)\sqrt{\wp\fil(x'';t)}\ket{\psi_x}\bra{\psi_{x''}}\delta_{x,x'}\delta_{x',x''}\\
&= \frac{1}{\wp(\fut{\bf O}_t|\past{\bf O}_t)} \sum_{x\in{\mathbb X}}E\rfil(x;t)\wp\fil(x;t)\ket{\psi_x}\bra{\psi_{x}}\\
& = \sum_{x\in{\mathbb X}} \wp\sm(x;t)\ket{\psi_x}\bra{\psi_x}\,.
\end{align}\\

\underline{\bf Criteria 4:} {\em \blk The smoothed quantum state must always be Hermitian and positive 
semidefinite, that is, it must be a $\mathfrak{S}$-class quantum state.}\\

Since both the filtered state and retrofiltered effect are always Hermitian and positive semidefinite, then it is trivial to show that $\rho\sm(t)$ is an $\mathfrak{S}$-class quantum state.

\subsection{Detailed derivation of Petz-Fuchs smoothed state in the Guevara-Wiseman scenario (\erf{eq:Petz-Wiseman_sm})}
In this section we will provide a more in-depth derivation of \erf{eq:Petz-Wiseman_sm}, \ie, the Petz-Fuchs smoothed state in the Guevara-Wiseman scenario. Let us begin with the dynamical equation for the true state. Since, in this scenario, we have two continuous-in-time measurements being performed on the system (one by Alice and the other by Bob), the unconditional dynamics is unraveled into $\hat{\cal E} = \sum_{\by_t,{\bf z}_t} \hat{\cal T}_{\by_t,{\bf z}_t}$, where $\by_t$ (${\bf z}_t$) denotes Alice's (Bob's) measurement outcomes at time $t$. The quantum trajectory formed by chaining these `true' conditional forward maps gives the (unnormalized) true state
\beq
\tilde{\rho}\god(t+\dd t) = \hat{\cal T}_{\by_t,{\bf z}_t}[\tilde{\rho}\god(t)]\,,
\eeq
where the norm of the state is $\Tr[\tilde{\rho}\god(t)] = \wp(\past{\bf O}_t,\past{\bf U}_t)$. However, in the Guevara-Wiseman scenario, Alice does not have knowledge of Bob's measurement outcome. As such, Alice's best description of the system at time $t$ given her observations up until then is to average over all measurement outcomes for Bob, \ie,
\beq
\tilde{\rho}\fil(t) = \sum_{{\bf z}_t} \tilde{\rho}\god(t)\,.
\eeq
It is easy to show \cite{GueWis15}, since $\hat{\cal E} = \sum_{\by_t}\hat{\cal F}_{\by_t}$, that Alice's best description of the system is identical to the standard filtered quantum state in \erf{q_fil}. 

To condition on the future measurement records, we apply the Petz recovery map, yielding
\begin{align}
\tilde{\rho}\god^{\rm PF}(t) &= {\cal R}^{\tilde{\rho}\god(t)}_{\by_t,{\bf z}_t}[\tilde{\rho}\god^{\rm PF}(t + \dd t)]\\
&= \sqrt{\tilde{\rho}\god(t)}\hat{E}\god(t)\sqrt{\tilde{\rho}\god(t)}\,,
\end{align}
where the second line follows from Sec.~\ref{App:derive_PF} and we have defined the `true' retrofiltered effect as $\hat{E}\god(t) = \hat{\cal T}\dg_{\by_t,{\bf z}_t}[\hat{E}\god(t + \dd t)]$ with final condition $\hat{E}\god(T)\propto \mathds{1}$. This state is conditioned on both Alice's and Bob's future measurement records, what we will call the Petz-Fuchs true state. However, Alice still does not know any of Bob's measurement outcomes, and similar to the filtering case, her description of the system is given by the average over all possible records $\both{\bf U}$ for Bob. That is,
\beq
\begin{split}
\tilde{\rho}\sm(t) &= \sum_{\both{\bf U}} \tilde{\rho}\god^{\rm PF}(t)\\
& = \sum_{\both{\bf U}} \sqrt{\tilde{\rho}\god(t)}\hat{E}\god(t)\sqrt{\tilde{\rho}\god(t)}\\
& = \sum_{\past{\bf U}_t} \sqrt{\tilde{\rho}\god(t)}\left[\sum_{\fut{\bf U}_t} \hat{E}\god(t)\right]\sqrt{\tilde{\rho}\god(t)}\,.
\end{split}
\eeq
Recognizing that $\hat{\cal F}_{\by_t} = \sum_{{\bf z}_t}\hat{\cal T}_{\by_t,{\bf z}_t}$, we have
\beq
\begin{split}
\tilde{\rho}\sm(t) & = \sum_{\past{\bf U}_t} \sqrt{\tilde{\rho}\god(t)}\hat{E}\rfil(t)\sqrt{\tilde{\rho}\god(t)}\\
&\propto \sum_{\past{\bf U}_t} \wp(\past{\bf U}_t|\past{\bf O}_t)\sqrt{\rho\god(t)}\hat{E}\rfil(t)\sqrt{\rho\god(t)}\,.
\end{split}
\eeq

\subsection{Alternative derivation of \erf{eq:Petz-Wiseman_sm} using classical registers}
In what follows, again, we provide a method of applying the Petz recovery map via CPTP maps to derive the smoothed state in the Guevara-Wiseman scenario, \ie, when the Kraus decomposition $\{K_\ell\}$ is fixed by Bob's measurement. In this scenario, there are two sets of measurement outcomes recorded by two sets of classical registers respectively. Let us begin with the initial quantum state $\rho_0$, from Alice's perspective, after performing measurements continuously until time $t$. The hybrid classical-quantum state is
\beq
\tilde{\varrho}\fil(t)=\sum_{\past{\bf U}_t}\hat{\cal F}_{\past{\bf O}_t,\past{\bf U}_t}[\rho_0]\otimes\op{\past{\bf O}_t}{\past{\bf O}_t}\otimes\op{\past{\bf U}_t}{\past{\bf U}_t},
\eeq
where $\hat{\cal F}_{\past{\bf O}_t,\past{\bf U}_t}$is the forward quantum operation conditioned on both Alice's and Bob's past measurement records from $t_0$ to $t$. Again, we working with unnormalized states here for simplicity.

We can describe the evolution of the continuous-in-time measurement process from time $t$ to $T$ by the CPTP map ${\cal E}_{t\to T}[\bullet] = \sum_{\fut{\bf O}_t,\fut{\bf U}_t}(\hat{\cal F}_{\fut{\bf O}_t,\fut{\bf U}_t}\otimes {\cal I}_{C_{\past{\bf O}_t,\past{\bf U}_t}})[\bullet]\otimes\op{\fut{\bf O}_t}{\fut{\bf O}_t}\otimes\op{\fut{\bf U}_t}{\fut{\bf U}_t}$, where $\hat{\cal F}_{\fut{\bf O}_t,\fut{\bf U}_t}$ is the forward quantum operation conditioned on all possible measurement outcomes from $t$ to $T$ that Alice and Bob could observe, and ${\cal I}_{C_{\past{\bf O}_t,\past{\bf U}_t}}$ is the identity channel on classical registers up to time $t$. Remember that, for maps that involve classical registers, we have omitted the $\hat{\bullet}$ accent to avoid confusion. This is the CPTP map we will use in the Petz recovery map to construct our retrodiction channel back to time $t$, \ie, ${\cal R}^{\tilde{\varrho}\fil(t)}_{{ \cal E}_{t\to T}}[\bullet]$. 

Again, we can now apply our retrodiction channel to the hybrid filtered state at the final time, $\tilde{\varrho}\fil(T)=\sum_{\both{\bf U}}\hat{\cal F}_{\both{\bf O},\both{\bf U}}[\rho_0]\otimes\op{\both{\bf O}}{\both{\bf O}}\otimes\op{\both{\bf U}}{\both{\bf U}}$, which contains all the information about which measurement record was realized, to obtain our smoothed state. That is 
\begin{equation}\label{seq:CPTP_Petz_2}
    \tilde{\varrho}\sm(t)={\cal R}^{\tilde{\varrho}\fil(t)}_{{ \cal E}_{t\to T}}[\tilde{\varrho}\fil(T)]=\sqrt{\tilde{\varrho}\fil(t)}{\cal E}^\dagger\left[\frac{1}{\sqrt{{\cal E}[\tilde{\varrho}\fil(t)]}}\tilde{\varrho}\fil(T)\frac{1}{\sqrt{{\cal E}[\tilde{\varrho}\fil(t)]}}\right]\sqrt{\tilde{\varrho}\fil(t)},
\end{equation}
Using the orthogonality of the classical registers, we have
\beq
\begin{split}
    \frac{1}{\sqrt{{\cal E}[\tilde{\varrho}\fil(t)]}}=&\sum_{\fut{\bf O}_t,\both{\bf U}}\frac{1}{\sqrt{\hat{\cal F}_{\both{\bf O},\both{\bf U}}[\rho_0]}}\otimes \op{\both{\bf O}}{\both{\bf O}}\otimes \op{\both{\bf U}}{\both{\bf U}},
\end{split}
\eeq
from which, we obtain
\beq
\begin{split}
    {\cal E}^\dagger\left[\frac{1}{\sqrt{{\cal E}[\tilde{\varrho}\fil(t)]}}\tilde{\varrho}\fil(T)\frac{1}{\sqrt{{\cal E}[\tilde{\varrho}\fil(t)]}}\right]
    &={\cal E}^\dagger[\mathds{1}_Q\otimes\op{\both{\bf O}}{\both{\bf O}}\otimes\mathds{1}_{C_{\both{\bf U}}}]\\
    &=\sum_{\fut{\bf U}_t}\hat{\cal F}^\dagger_{\fut{\bf O}_t,\fut{\bf U}_t}[\mathds{1}_Q]\otimes\op{\past{\bf O}_t}{\past{\bf O}_t}\otimes \mathds{1}_{C_{{\past{\bf U}}_t}}\\
    &=\hat{E}\rfil(t)\otimes\op{\past{\bf O}_t}{\past{\bf O}_t}\otimes \mathds{1}_{C_{{\past{\bf U}}_t}},
\end{split}
\eeq
where $\mathds{1}_{C_{{\past{\bf U}}_t}}$ denotes the identity on the classical registers for $\past{\bf U}_t$.
Substituting this into \erf{seq:CPTP_Petz_2} yields 
\beq
\begin{split}
    \tilde{\varrho}\sm(t)=&\sqrt{\tilde{\varrho}\fil(t)}(\hat{E}\rfil(t)\otimes\op{\past{\bf O}_t}{\past{\bf O}_t}\otimes \mathds{1}_{C_{{\past{\bf U}}_t}})\sqrt{\tilde{\varrho}\fil(t)}\\
    =&\sum_{\past{\bf U}_t}\sqrt{\hat{\cal F}_{\past{\bf O}_t,\past{\bf U}_t}[\rho_0]}\hat{E}\rfil(t)\sqrt{\hat{\cal F}_{\past{\bf O}_t,\past{\bf U}_t}[\rho_0]}\otimes\op{\past{\bf O}_t}{\past{\bf O}_t}\otimes\op{\past{\bf U}_t}{\past{\bf U}_t}.
\end{split}
\eeq
Finally, tracing out the classical system, we obtain
\beq
\begin{split}
\tilde{\rho}\sm(t)&=\Tr_C[\tilde{\varrho}\sm(t)]\\
&=\sum_{\past{\bf U}_t}\sqrt{\tilde{\rho}\god(t)}\hat{E}\rfil(t)\sqrt{\tilde{\rho}\god(t)}\\
&= \sum_{\past{\bf U}_t} \wp(\past{\bf O}_t,\past{\bf U}_t) \sqrt{\rho\god(t)}\hat{E}\rfil(t)\sqrt{\rho\god(t)}\\
&\propto \sum_{\past{\bf U}_t} \wp(\past{\bf U}_t|\past{\bf O}_t) \sqrt{\rho\god(t)}\hat{E}\rfil(t)\sqrt{\rho\god(t)}\,,
\end{split}
\eeq
where $\hat{\cal F}_{\past{\bf O}_t,\past{\bf U}_t}[\rho_0] = \tilde{\rho}\god(t)$ and, in the final line, since the $\past{\bf O}_t$ is fixed, $\wp(\past{\bf O}_t)$ is a constant, $\wp(\past{\bf O}_t,\past{\bf U}_t) \propto \wp(\past{\bf U}_t|\past{\bf O}_t)$.

Additionally, by normalizing the state, and noting that $\wp(\fut{\bf O}_t|\past{\bf O}_t,\past{\bf U}_t) = \Tr[\hat{E}\rfil(t) \rho\god(t)]$, we can rewrite this smoothed state as 
\beq
\begin{split}
    \rho\sm(t) &= \frac{\sum_{\past{\bf U}_t} \wp(\past{\bf U}_t|\past{\bf O}_t) \sqrt{\rho\god(t)}\hat{E}\rfil(t)\sqrt{\rho\god(t)}}{\Tr\left[\sum_{\past{\bf U}_t} \wp(\past{\bf U}_t|\past{\bf O}_t) \sqrt{\rho\god(t)}\hat{E}\rfil(t)\sqrt{\rho\god(t)}\right]}\\
    &= \frac{1}{\sum_{\past{\bf U}_t} \wp(\past{\bf U}_t|\past{\bf O}_t) \wp(\fut{\bf O}_t|\past{\bf O}_t,\past{\bf U}_t)} \sum_{\past{\bf U}_t} \wp(\past{\bf U}_t|\past{\bf O}_t)\wp(\fut{\bf O}_t|\past{\bf O}_t,\past{\bf U}_t) \frac{\sqrt{\rho\god(t)}\hat{E}\rfil(t)\sqrt{\rho\god(t)}}{\Tr\left[\sqrt{\rho\god(t)}\hat{E}\rfil(t)\sqrt{\rho\god(t)}\right]}\\
    &= \sum_{\past{\bf U}_t} \wp(\past{\bf U}_t|\both{\bf O}) \frac{\sqrt{\rho\god(t)}\hat{E}\rfil(t)\sqrt{\rho\god(t)}}{\Tr\left[\sqrt{\rho\god(t)}\hat{E}\rfil(t)\sqrt{\rho\god(t)}\right]}\\
    &= \sum_{\past{\bf U}_t}\wp(\past{\bf U}_t|\both{\bf O})\hat{\cal R}_{\fut{\bf O}_t}^{\rho\god(t)}[\rho_{\both{\bf O},\past{\bf U}_t}(T)],
    \end{split}
\end{equation}
where $\hat{\cal R}_{\fut{\bf O}_t}^{\rho\god(t)}$ is the Petz recovery map of forward quantum operation $\hat{\cal F}_{\fut{\bf O}_t}$ with reference prior ${\rho\god(t)}$ and $\rho_{\both{\bf O},\past{\bf U}_t}(T) = \hat{\cal F}_{\fut{\bf O}_t}[\tilde{\rho}\god(t)]/\Tr\left[\hat{\cal F}_{\fut{\bf O}_t}[\tilde{\rho}\god(t)]\right]$.

Comparing this result with the definition of the smoothed state in Ref.~\cite{GueWis15}, $ {\rho}\sm^{\rm GW}(t) = \sum_{\past{\bf U}_t} \wp(\past{\bf U}_t|\both{\bf O})\rho\god(t)$, we see that both compute the average at time $t$ via same probability distribution, \ie, $\wp(\past{\bf U}_t|\both{\bf O})$. The difference is that the quantum states that are being averaged are different. Now, if all the true states are pure, our retrodictive approach to smoothing and the Guevara-Wiseman approach are identical. This is because the conditional retrodiction map cannot improve upon the pure state $\rho\god(t)$ known at time $t$. However, when the true state is impure, they are not the same in general and we see that the Guevara-Wiseman approach is conditioning on the future information differently. Note, we can also expect in this latter case that the resulting smoothed state will, on average, be purer than Guevara and Wiseman's state, since $\hat{\cal R}_{\fut{\bf O}_t}^{\rho\god(t)}[\rho_{\both{\bf O},\past{\bf U}_t}(T)]$ is generally purer than the corresponding filtered state, $\rho\god(t)$. However, it should be stressed that the optimality \cite{LGW-PRA21} of our retrodictive state, in this case, is unknown.

\subsection{Single trajectory comparison to other smoothing approaches}
In this section, we compare a single trajectory of the Petz-Fuchs smoothed state (red dashed), the smoothed weak-valued state (green), defined following the notation of Ref.~\cite{LCW19} as
\beq
\varrho\swv(t) \propto \rho\fil(t) \hat{E}\rfil(t) + \hat{E}\rfil(t) \rho\fil(t)\,,
\eeq
where in this section $\varrho$ does {\em not} denote hybrid quantum-classical state, and the Guevara-Wiseman smoothed state. See Fig.~\ref{Sfig:Trajectories}. The scenario we consider is the same as in Fig.~\ref{fig:Trajectories} of the main text, where, in the case of Guevara-Wiseman smoothing, we consider all 3 different unravelings for Bob's measurement: photon detection (magenta), X-homodyne (cyan) and Y-homodyne (orange). Note, again, $\ex{\s{y}} = 0$ in all cases. 

Qualitatively, in terms of the trajectories, all the smoothed states look similar, with the only major difference being the degree of purification of the state. The greatest similarity between states is seen between the Petz-Fuchs state and the smoothed weak-valued state, where inbetween the initial time and the jump there is very little difference. This is due to the non-commutativity of the filtered state and retrofiltered effect being small enough that $\varrho\swv\approx\rho\sm^{{\rm PF}}$, where the superscript here denotes the Petz-Fuchs smoothed state. To see this approximation, consider
\beq
\begin{split}
\rho\sm^{\rm PF}(t) &= \frac{\sqrt{\rho\fil(t)}\hat{E}\rfil(t)\sqrt{\rho\fil(t)}}{\Tr[\rho\fil(t)\hat{E}\rfil(t)]}\\
& =\frac{1}{2{\Tr[\rho\fil(t)\hat{E}\rfil(t)]}}\left(\sqrt{\rho\fil(t)}\hat{E}\rfil(t)\sqrt{\rho\fil(t)} + \sqrt{\rho\fil(t)}\hat{E}\rfil(t)\sqrt{\rho\fil(t)}\right)\\
& = \frac{1}{2{\Tr[\rho\fil(t)\hat{E}\rfil(t)]}}\left(\hat{E}\rfil\rho\fil(t) + \rho\fil(t)\hat{E}\rfil(t) + \left[\left[\hat{E}\rfil(t),\sqrt{\rho\fil(t)}\right],\sqrt{\rho\fil(t)}\right]\right)\\
& =  \varrho\swv(t) + \frac{\left[\left[\hat{E}\rfil(t),\sqrt{\rho\fil(t)}\right],\sqrt{\rho\fil(t)}\right]}{\Tr[\rho\fil(t)\hat{E}\rfil(t)]}\,.
\end{split}
\eeq
Thus, when $\left[\hat{E}\rfil(t),\sqrt{\rho\fil(t)}\right]$ is small, we get $\varrho\swv\approx\rho\sm^{{\rm PF}}$. However, when the commutativity large, as is the case near the jump, we see that the smoothed weak-valued state becomes unphysical ($P>1$). This relationship between the Petz-Fuchs smoothed state and the smoothed weak-valued state is unsurprising as they both belong to the class of symmetrized products 
\beq
\frac{1}{2\Tr[\rho\hat{E}]}\left(\rho^{\alpha}\hat{E}\rho^{1 - \alpha} + \rho^{1 - \alpha}\hat{E}\rho^{\alpha}\right)\,,
\eeq
where $\alpha \in [1/2,1]$. As an aside, it would be an interesting to see whether the parameter $\alpha$ can be thought of as a measure of ``classicality'', since in the classical regime, all of these products are identical and yield a physical state, however, when we have some degree of non-commutativity, they will be different and some unphysical.

Comparing now, the Petz-Fuchs smoothed state to the Guevara-Wiseman smoothed states, we can see that former is always purer than the latter, for any of the three choices of measurement by Bob. This should not be surprising, however, since Guevara and Wiseman's smoothed state was not designed to maximize the purity of the state, but rather minimize, on average, the Hilbert-Schmidt distance (or relative entropy) to the true state \cite{LGW-PRA21}. One can also understand the differences in the purity between the three cases of Guevara-Wiseman smoothing based on the first order correlations in the measurement currents \cite{CGW19}.

\begin{figure}
    \centering
    \includegraphics[width=\linewidth]{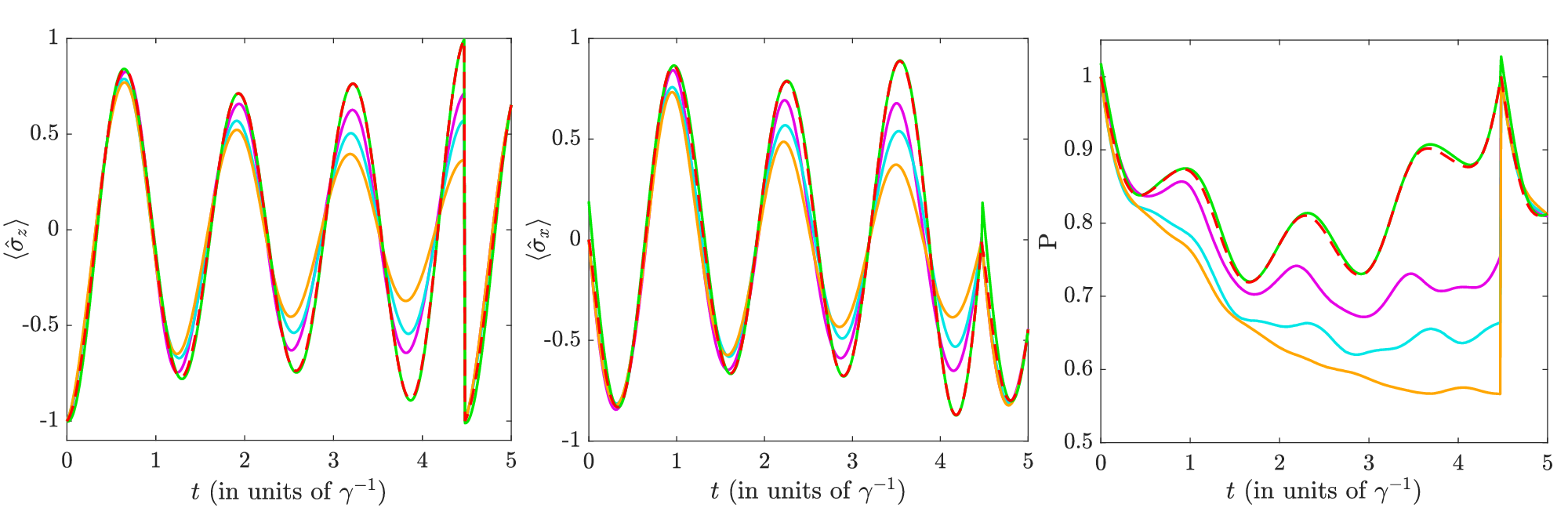}
    \caption{A sample trajectory of different smoothed states in terms of their Bloch components as well as their corresponding purity. For reference, the red dashed line shows the trajectory of Petz-Fuchs smoothed state. The Guevara-Wiseman smoothed states are denoted by the magenta, cyan and orange lines for photon counting, X-Homodyne and Y-Homodyne detection unraveling respectively. Lastly, the green line is for the smoothed weak-valued state. Note, in this case $\ex{\s{y}}(t) = 0$ for all five states. For this trajectory, only a single photon is detected during the evolution, at $t\approx4.5\gamma\inv$. Here, $\Omega=5\gamma$, $\bar{n} = 0.5$.}
    \label{Sfig:Trajectories}
\end{figure}

\subsection{Numerics}
The numerics that appear in this paper are largely described in Refs.~\cite{GueWis20,CGW19} and the references therein. However, for the sake of completeness, we will present the necessary theory required to reproduce the results presented. We begin by considering the conditional dynamics of the optical Bloch equations. To numerically compute the filtered state, we begin by discretizing the dynamics into the following
\beq\label{eq:disc_filt}
\tilde\rho\fil(t + \Delta t) = {\grn\hat{U}(t+\Delta t,t)} {\blk \hat{M}_{\by_t}}{\blu \sum_{\ell} \hat{K}_\ell}\tilde\rho\fil(t) {\blu\hat{K}\dg_\ell} {\blk \hat{M}\dg_{\by_t}}{\grn\hat{U}\dg(t+\Delta t,t)}.
\eeq
That is, we break the evolution into three parts: the unitary evolution (green), the measurement (red) and the decoherence (blue). Beginning with the unitary dynamics, this term encompasses the dynamics the qubit experiences due to the classical driving. Since the driving Hamiltonian of the system is time independent, we have ${\grn \hat{U}(t+\dd t,t)} = {\rm exp}(-i\hat{H}\Delta t)$. 

For the measurement dynamics, we consider two types of measurement, photon detection and homodyne detection. In either case, we first must to generate a measurement record. As we are following Ref.~\cite{GueWis20}, the measurement operators are described with respect to an {\em ostensible} distribution $\wp\ost(\by_t)$. The ostensible distribution is an arbitrarily chosen distribution, so long as it does not prevent renormalization, and is used to generate the measurement record (be it the photon detection time or the homodyne photocurrent). Naturally, to to ensure that states generated with these ostensible records are unbiased, \ie, so that ensemble averages performed with these records converge to the correct answer, the quantum maps need to be adjusted accordingly. In particular, for a quantum jump unraveling of the dynamics, the adjusted measurement operators are
\beq
{\blk \hat{M}_{\by_t = 1}} = \sqrt{\frac{\Delta t}{\wp\ost(\by_t)}}\hat{c}\,,\qquad{\rm and}\qquad {\blk \hat{M}_{\bf y_t = 0}} = \sqrt{\frac{\mathds{1} - \hat{c}\dg\hat{c}\Delta t}{1 - \wp\ost(\by_t)}}\approx \mathds{1} - \frac{1}{2}(\hat{c}\dg \hat{c}\Delta t - \wp\ost(\by_t)) - \frac{1}{8}(\hat{c}\dg\hat{c}\Delta t - \wp\ost(\by_t))^2\,,
\eeq
where $\hat{c}$ is the measured Lindblad operator (in our case $\hat{c} = \sqrt{\gamma (\bar{n} + 1)}\s{-}$) and we have kept terms up to $\Delta t^2$ for greater accuracy. In our case, the ostensible distribution we choose is the actual jump probability $\wp(\by_t = 1) = \Delta t\Tr[\hat{c}\dg\hat{c} {\blu\sum_{\ell} \hat{K}_\ell}\tilde\rho\fil(t) {\blu\hat{K}\dg_\ell}]/\Tr[{\blu \sum_\ell\hat{K}_\ell}\tilde\rho\fil(t) {\blu\hat{K}\dg_\ell}]$. Note, in this case, this amounts to normalizing the filtered state in \erf{eq:disc_filt}. This is not problematic for computing the Petz-Fuchs smoothed state in \erf{eq:q_sm}, as any proportionalities (like the normalization of the filtered state and similarly the retrofiltered effect) can be corrected by renormalizing the smoothed state. 

In the case of a homodyne unraveling, the ostensible distribution for the homodyne current $\by_t$ is taken to be a zero mean Guassian with variance = $\Delta t$, \ie, $\wp\ost(\by_t) = g(\by_t;0,\Delta t)$. The measurement operator is \cite{GueWis20} 
\beq\label{eq:M_hom}
{\blk \hat{M}_{\by_t}} = \mathds{1} + \hat{c}e^{i\phi}\by_t\Delta t - \frac{1}{2}\hat{c}\dg\hat{c}\Delta t + \frac{1}{8}(\hat{c}\dg\hat{c}\Delta t)^2\,,
\eeq
where $\phi$ is the homodyne phase. For an X-homodyne measurement, $\phi = 0$ and for a Y-homodyne, $\phi = \pi/2$. However, for a faster convergence when performing ensemble averages, we compute the actual photocurrent by shifting the ostensibly generated photocurrent (denoted here by $\dd W$) to 
\beq
\by_t = \frac{\Tr[(\hat{c}e^{i\phi} + \hat{c}\dg e^{-i\phi}){\blu\sum_{\ell} \hat{K}_\ell}\tilde\rho\fil(t) {\blu\hat{K}\dg_\ell}]}{\Tr\left[{\blu \sum_\ell\hat{K}_\ell}\tilde\rho\fil(t) {\blu\hat{K}\dg_\ell}\right]} + \dd W\,,
\eeq
before applying \erf{eq:M_hom} and renormalizing.  

Finally, onto the decoherence channel. This part of the dynamics captures the remaining environmental interactions that are not captured by the measurement. For generality, we have been expressing this decoherence channel in terms of its Kraus decomposition \cite{Kraus83} with Kraus operators $\hat{K}_\ell$ chosen such that they reproduce remaining, unmeasured dissipator in the continuous-time-limit. As in our case, the choice of this Kraus operator has no baring on the resulting smoothed state, we have chosen the simplest Kraus operators that reproduces a dissipator ${\cal D}[\hat{a}]\bullet$, 
\beq
{\blu \hat{K}_1} = \hat{a}\sqrt{\Delta t}\,,\qquad {\rm and}\qquad {\blu \hat{K}_0} = \sqrt{1 - \hat{a}\dg\hat{a}\Delta t} \approx  1 - \frac{1}{2} \hat{a}\dg\hat{a}\Delta t - \frac{1}{8}(\hat{a}\dg\hat{a}\Delta t)^2\,,
\eeq
where, once again, we have truncated to second order in $\Delta t$, for increased accuracy. In the example we consider, $\hat{a} = \sqrt{\gamma\bar{n}}\s{+}$.

To ensure that we are correctly generating the trajectories and that our ensemble averages are unbiased, for each measurement setting, we have plotted the ensemble average of the filtered (blue) and smoothed (red) states over 30000 trajectories. We can see in Fig.~\ref{Sfig:Avg_Trajectories} that the both converge to the unconditioned solution (black), as one would expect, with an error smaller than the width of the line. 

\begin{figure}[h]
\includegraphics[scale=0.35]{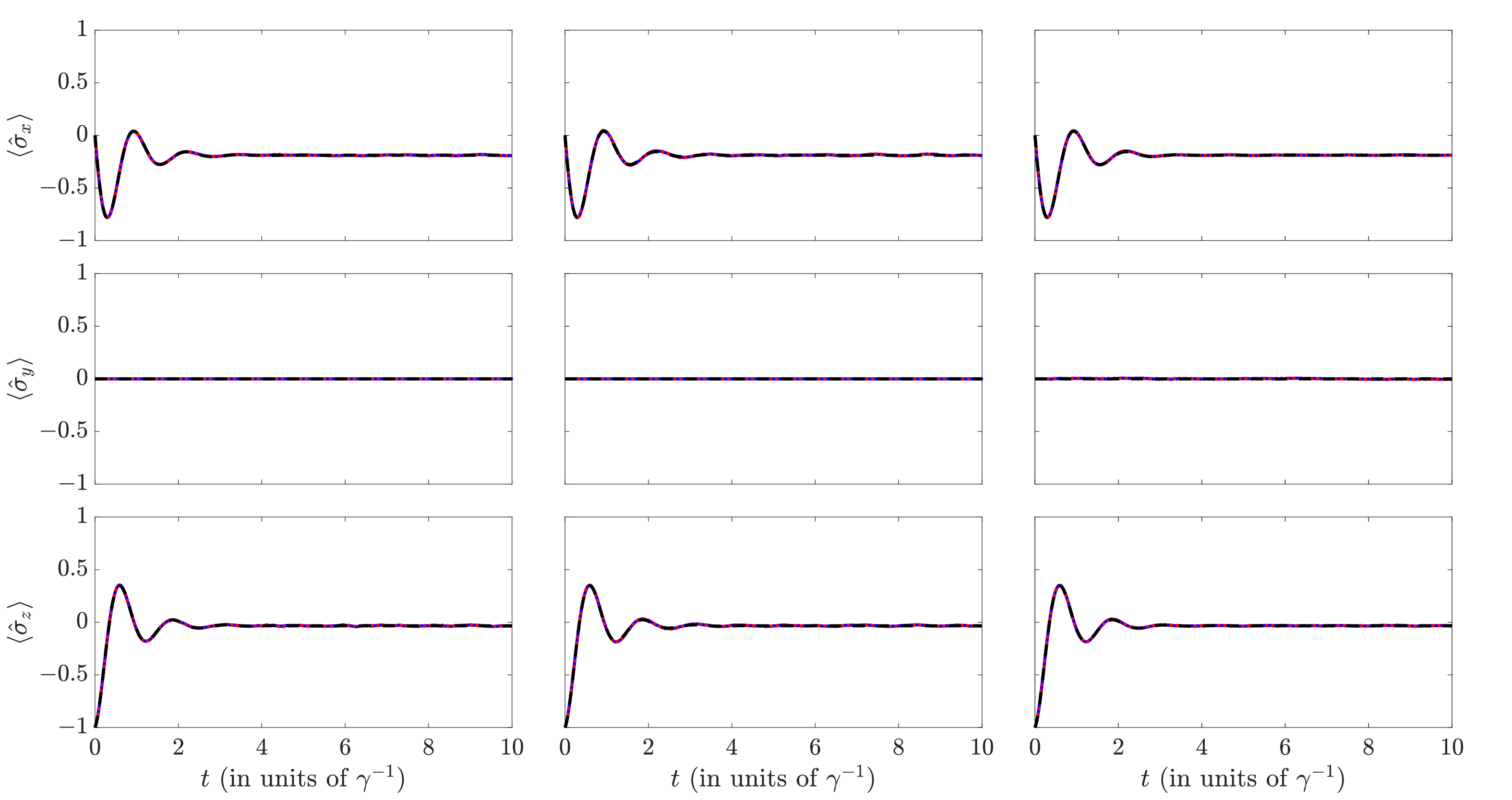}
\caption{The ensemble average of the filtered (blue solid line) and smoothed (red dotted line) states for photon detection (left column), X-homodyne (center column) and Y-homodyne measurements (right column). The ensemble average for either the filtered or smoothed state should recover the unconditioned dynamics (black dashed line), for which we see a great agreement with. For each detection method, we have plotted the $x$-, $y$- and $z$-components of the Bloch vector for each state. Each measurement case has been generated using 30000 random trajectories. Here, $\Omega=5\gamma$, $\bar{n} = 0.5$.}
\label{Sfig:Avg_Trajectories}
\end{figure}

\end{widetext}


\begin{thebibliography}{56}%
\makeatletter
\providecommand \@ifxundefined [1]{%
 \@ifx{#1\undefined}
}%
\providecommand \@ifnum [1]{%
 \ifnum #1\expandafter \@firstoftwo
 \else \expandafter \@secondoftwo
 \fi
}%
\providecommand \@ifx [1]{%
 \ifx #1\expandafter \@firstoftwo
 \else \expandafter \@secondoftwo
 \fi
}%
\providecommand \natexlab [1]{#1}%
\providecommand \enquote  [1]{``#1''}%
\providecommand \bibnamefont  [1]{#1}%
\providecommand \bibfnamefont [1]{#1}%
\providecommand \citenamefont [1]{#1}%
\providecommand \href@noop [0]{\@secondoftwo}%
\providecommand \href [0]{\begingroup \@sanitize@url \@href}%
\providecommand \@href[1]{\@@startlink{#1}\@@href}%
\providecommand \@@href[1]{\endgroup#1\@@endlink}%
\providecommand \@sanitize@url [0]{\catcode `\\12\catcode `\$12\catcode
  `\&12\catcode `\#12\catcode `\^12\catcode `\_12\catcode `\%12\relax}%
\providecommand \@@startlink[1]{}%
\providecommand \@@endlink[0]{}%
\providecommand \url  [0]{\begingroup\@sanitize@url \@url }%
\providecommand \@url [1]{\endgroup\@href {#1}{\urlprefix }}%
\providecommand \urlprefix  [0]{URL }%
\providecommand \Eprint [0]{\href }%
\providecommand \doibase [0]{http://dx.doi.org/}%
\providecommand \selectlanguage [0]{\@gobble}%
\providecommand \bibinfo  [0]{\@secondoftwo}%
\providecommand \bibfield  [0]{\@secondoftwo}%
\providecommand \translation [1]{[#1]}%
\providecommand \BibitemOpen [0]{}%
\providecommand \bibitemStop [0]{}%
\providecommand \bibitemNoStop [0]{.\EOS\space}%
\providecommand \EOS [0]{\spacefactor3000\relax}%
\providecommand \BibitemShut  [1]{\csname bibitem#1\endcsname}%
\let\auto@bib@innerbib\@empty
\bibitem [{\citenamefont {Rauch}(1963)}]{Rauch63}%
  \BibitemOpen
  \bibfield  {author} {\bibinfo {author} {\bibfnamefont {H.~E.}\ \bibnamefont
  {Rauch}},\ }\bibfield  {title} {\enquote {\bibinfo {title} {Solutions to the
  linear smoothing problem},}\ }\href {\doibase 10.1109/TAC.1963.1105600}
  {\bibfield  {journal} {\bibinfo  {journal} {IEEE Trans. Automat. Contr.}\
  }\textbf {\bibinfo {volume} {8}},\ \bibinfo {pages} {371--372} (\bibinfo
  {year} {1963})}\BibitemShut {NoStop}%
\bibitem [{\citenamefont {Fraser}\ and\ \citenamefont
  {Potter}(1969)}]{FraPot69}%
  \BibitemOpen
  \bibfield  {author} {\bibinfo {author} {\bibfnamefont {D.~C.}\ \bibnamefont
  {Fraser}}\ and\ \bibinfo {author} {\bibfnamefont {J.}~\bibnamefont
  {Potter}},\ }\bibfield  {title} {\enquote {\bibinfo {title} {The optimum
  linear smoother as a combination of two optimum linear filters},}\ }\href
  {\doibase 10.1109/TAC.1969.1099196} {\bibfield  {journal} {\bibinfo
  {journal} {IEEE Trans. Automat. Contr.}\ }\textbf {\bibinfo {volume} {14}},\
  \bibinfo {pages} {387--390} (\bibinfo {year} {1969})}\BibitemShut {NoStop}%
\bibitem [{\citenamefont {Weinert}(2001)}]{Weinert01}%
  \BibitemOpen
  \bibfield  {author} {\bibinfo {author} {\bibfnamefont {H.~L.}\ \bibnamefont
  {Weinert}},\ }\href@noop {} {\emph {\bibinfo {title} {Fixed Interval
  Smoothing for State Space Models}}}\ (\bibinfo  {publisher} {Kluwer
  Academic},\ \bibinfo {address} {New York},\ \bibinfo {year}
  {2001})\BibitemShut {NoStop}%
\bibitem [{\citenamefont {S{\"a}rkk{\"a}}(2013)}]{Sarkka13}%
  \BibitemOpen
  \bibfield  {author} {\bibinfo {author} {\bibfnamefont {S.}~\bibnamefont
  {S{\"a}rkk{\"a}}},\ }\href@noop {} {\emph {\bibinfo {title} {Bayesian
  filtering and smoothing}}},\ Vol.~\bibinfo {volume} {3}\ (\bibinfo
  {publisher} {Cambridge University Press},\ \bibinfo {address} {Cambridge,
  England},\ \bibinfo {year} {2013})\BibitemShut {NoStop}%
\bibitem [{\citenamefont {Kalman}(1960)}]{Kalman60}%
  \BibitemOpen
  \bibfield  {author} {\bibinfo {author} {\bibfnamefont {R.~E.}\ \bibnamefont
  {Kalman}},\ }\bibfield  {title} {\enquote {\bibinfo {title} {A new approach
  to linear filtering and prediction problems},}\ }\href {\doibase
  10.1115/1.3662552} {\bibfield  {journal} {\bibinfo  {journal} {J. Basic
  Eng.}\ }\textbf {\bibinfo {volume} {82}},\ \bibinfo {pages} {35--45}
  (\bibinfo {year} {1960})}\BibitemShut {NoStop}%
\bibitem [{\citenamefont {Kushner}(1964)}]{Kushner64}%
  \BibitemOpen
  \bibfield  {author} {\bibinfo {author} {\bibfnamefont {H.~J.}\ \bibnamefont
  {Kushner}},\ }\bibfield  {title} {\enquote {\bibinfo {title} {{On the
  differential equations satisfied by conditional probability densities of
  Markov processes, with applications}},}\ }\href {\doibase 10.1137/0302009}
  {\bibfield  {journal} {\bibinfo  {journal} {J. Soc. Ind. Appl. Math. A}\
  }\textbf {\bibinfo {volume} {2}},\ \bibinfo {pages} {106--119} (\bibinfo
  {year} {1964})}\BibitemShut {NoStop}%
\bibitem [{\citenamefont {Jacobs}(1996)}]{Jac96}%
  \BibitemOpen
  \bibfield  {author} {\bibinfo {author} {\bibfnamefont {O.~L.~R.}\
  \bibnamefont {Jacobs}},\ }\href@noop {} {\emph {\bibinfo {title}
  {Introduction to Control Theory}}},\ \bibinfo {edition} {2nd}\ ed.\ (\bibinfo
   {publisher} {Oxford University Press},\ \bibinfo {address} {New York},\
  \bibinfo {year} {1996})\BibitemShut {NoStop}%
\bibitem [{\citenamefont {Jazwinski}(2007)}]{Jaz07}%
  \BibitemOpen
  \bibfield  {author} {\bibinfo {author} {\bibfnamefont {A.~H.}\ \bibnamefont
  {Jazwinski}},\ }\href@noop {} {\emph {\bibinfo {title} {Stochastic processes
  and filtering theory}}}\ (\bibinfo  {publisher} {Dover Publications},\
  \bibinfo {address} {New York},\ \bibinfo {year} {2007})\BibitemShut {NoStop}%
\bibitem [{\citenamefont {Tsang}(2009{\natexlab{a}})}]{Tsa-PRL09}%
  \BibitemOpen
  \bibfield  {author} {\bibinfo {author} {\bibfnamefont {M.}~\bibnamefont
  {Tsang}},\ }\bibfield  {title} {\enquote {\bibinfo {title} {Time-symmetric
  quantum theory of smoothing},}\ }\href {\doibase
  10.1103/PhysRevLett.102.250403} {\bibfield  {journal} {\bibinfo  {journal}
  {Phys. Rev. Lett.}\ }\textbf {\bibinfo {volume} {102}},\ \bibinfo {pages}
  {250403} (\bibinfo {year} {2009}{\natexlab{a}})}\BibitemShut {NoStop}%
\bibitem [{\citenamefont {Tsang}(2009{\natexlab{b}})}]{Tsa-PRA09}%
  \BibitemOpen
  \bibfield  {author} {\bibinfo {author} {\bibfnamefont {M.}~\bibnamefont
  {Tsang}},\ }\bibfield  {title} {\enquote {\bibinfo {title} {Optimal waveform
  estimation for classical and quantum systems via time-symmetric smoothing},}\
  }\href {\doibase 10.1103/PhysRevA.80.033840} {\bibfield  {journal} {\bibinfo
  {journal} {Phys. Rev. A}\ }\textbf {\bibinfo {volume} {80}},\ \bibinfo
  {pages} {033840} (\bibinfo {year} {2009}{\natexlab{b}})}\BibitemShut
  {NoStop}%
\bibitem [{\citenamefont {Chantasri}\ \emph {et~al.}(2013)\citenamefont
  {Chantasri}, \citenamefont {Dressel},\ and\ \citenamefont {Jordan}}]{CDJ13}%
  \BibitemOpen
  \bibfield  {author} {\bibinfo {author} {\bibfnamefont {A.}~\bibnamefont
  {Chantasri}}, \bibinfo {author} {\bibfnamefont {J.}~\bibnamefont {Dressel}},
  \ and\ \bibinfo {author} {\bibfnamefont {A.~N.}\ \bibnamefont {Jordan}},\
  }\bibfield  {title} {\enquote {\bibinfo {title} {Action principle for
  continuous quantum measurement},}\ }\href {\doibase
  10.1103/PhysRevA.88.042110} {\bibfield  {journal} {\bibinfo  {journal} {Phys.
  Rev. A}\ }\textbf {\bibinfo {volume} {88}},\ \bibinfo {pages} {042110}
  (\bibinfo {year} {2013})}\BibitemShut {NoStop}%
\bibitem [{\citenamefont {Weber}\ \emph {et~al.}(2014)\citenamefont {Weber},
  \citenamefont {Chantasri}, \citenamefont {Dressel}, \citenamefont {Jordan},
  \citenamefont {Murch},\ and\ \citenamefont {Siddiqi}}]{Web14}%
  \BibitemOpen
  \bibfield  {author} {\bibinfo {author} {\bibfnamefont {S.~J.}\ \bibnamefont
  {Weber}}, \bibinfo {author} {\bibfnamefont {A.}~\bibnamefont {Chantasri}},
  \bibinfo {author} {\bibfnamefont {J.}~\bibnamefont {Dressel}}, \bibinfo
  {author} {\bibfnamefont {A.~N.}\ \bibnamefont {Jordan}}, \bibinfo {author}
  {\bibfnamefont {K.~W.}\ \bibnamefont {Murch}}, \ and\ \bibinfo {author}
  {\bibfnamefont {I.}~\bibnamefont {Siddiqi}},\ }\bibfield  {title} {\enquote
  {\bibinfo {title} {Mapping the optimal route between two quantum states},}\
  }\href {\doibase 10.1038/nature13559} {\bibfield  {journal} {\bibinfo
  {journal} {Nature}\ }\textbf {\bibinfo {volume} {511}},\ \bibinfo {pages}
  {570--573} (\bibinfo {year} {2014})}\BibitemShut {NoStop}%
\bibitem [{\citenamefont {Gammelmark}\ \emph {et~al.}(2013)\citenamefont
  {Gammelmark}, \citenamefont {Julsgaard},\ and\ \citenamefont
  {M\o{}lmer}}]{GJM13}%
  \BibitemOpen
  \bibfield  {author} {\bibinfo {author} {\bibfnamefont {S.}~\bibnamefont
  {Gammelmark}}, \bibinfo {author} {\bibfnamefont {B.}~\bibnamefont
  {Julsgaard}}, \ and\ \bibinfo {author} {\bibfnamefont {K.}~\bibnamefont
  {M\o{}lmer}},\ }\bibfield  {title} {\enquote {\bibinfo {title} {{Past quantum
  states of a monitored system}},}\ }\href {\doibase
  10.1103/PhysRevLett.111.160401} {\bibfield  {journal} {\bibinfo  {journal}
  {Phys. Rev. Lett.}\ }\textbf {\bibinfo {volume} {{\bf 111}}},\ \bibinfo
  {pages} {160401} (\bibinfo {year} {2013})}\BibitemShut {NoStop}%
\bibitem [{\citenamefont {Gammelmark}\ \emph {et~al.}(2014)\citenamefont
  {Gammelmark}, \citenamefont {M{\o}lmer}, \citenamefont {Alt}, \citenamefont
  {Kampschulte},\ and\ \citenamefont {Meschede}}]{GamMol14}%
  \BibitemOpen
  \bibfield  {author} {\bibinfo {author} {\bibfnamefont {S.}~\bibnamefont
  {Gammelmark}}, \bibinfo {author} {\bibfnamefont {K.}~\bibnamefont
  {M{\o}lmer}}, \bibinfo {author} {\bibfnamefont {W.}~\bibnamefont {Alt}},
  \bibinfo {author} {\bibfnamefont {T.}~\bibnamefont {Kampschulte}}, \ and\
  \bibinfo {author} {\bibfnamefont {D.}~\bibnamefont {Meschede}},\ }\bibfield
  {title} {\enquote {\bibinfo {title} {{Hidden Markov model of atomic quantum
  jump dynamics in an optically probed cavity}},}\ }\href {\doibase
  10.1103/PhysRevA.89.043839} {\bibfield  {journal} {\bibinfo  {journal} {Phys.
  Rev. A}\ }\textbf {\bibinfo {volume} {89}},\ \bibinfo {pages} {043839}
  (\bibinfo {year} {2014})}\BibitemShut {NoStop}%
\bibitem [{\citenamefont {Guevara}\ and\ \citenamefont
  {Wiseman}(2015)}]{GueWis15}%
  \BibitemOpen
  \bibfield  {author} {\bibinfo {author} {\bibfnamefont {I.}~\bibnamefont
  {Guevara}}\ and\ \bibinfo {author} {\bibfnamefont {H.~M.}\ \bibnamefont
  {Wiseman}},\ }\bibfield  {title} {\enquote {\bibinfo {title} {Quantum state
  smoothing},}\ }\href {\doibase 10.1103/PhysRevLett.115.180407} {\bibfield
  {journal} {\bibinfo  {journal} {Phys. Rev. Lett.}\ }\textbf {\bibinfo
  {volume} {115}},\ \bibinfo {pages} {180407} (\bibinfo {year}
  {2015})}\BibitemShut {NoStop}%
\bibitem [{\citenamefont {Budini}(2017)}]{Budini17}%
  \BibitemOpen
  \bibfield  {author} {\bibinfo {author} {\bibfnamefont {A.~A.}\ \bibnamefont
  {Budini}},\ }\bibfield  {title} {\enquote {\bibinfo {title} {Smoothed
  quantum-classical states in time-irreversible hybrid dynamics},}\ }\href
  {\doibase 10.1103/PhysRevA.96.032118} {\bibfield  {journal} {\bibinfo
  {journal} {Phys. Rev. A}\ }\textbf {\bibinfo {volume} {96}},\ \bibinfo
  {pages} {032118} (\bibinfo {year} {2017})}\BibitemShut {NoStop}%
\bibitem [{\citenamefont {Chantasri}\ \emph {et~al.}(2021)\citenamefont
  {Chantasri}, \citenamefont {Guevara}, \citenamefont {Laverick},\ and\
  \citenamefont {Wiseman}}]{CGLW21}%
  \BibitemOpen
  \bibfield  {author} {\bibinfo {author} {\bibfnamefont {A.}~\bibnamefont
  {Chantasri}}, \bibinfo {author} {\bibfnamefont {I.}~\bibnamefont {Guevara}},
  \bibinfo {author} {\bibfnamefont {K.~T.}\ \bibnamefont {Laverick}}, \ and\
  \bibinfo {author} {\bibfnamefont {H.~M.}\ \bibnamefont {Wiseman}},\
  }\bibfield  {title} {\enquote {\bibinfo {title} {Unifying theory of quantum
  state estimation using past and future information},}\ }\href {\doibase
  10.1016/j.physrep.2021.07.003} {\bibfield  {journal} {\bibinfo  {journal}
  {Phys. Rep.}\ }\textbf {\bibinfo {volume} {930}},\ \bibinfo {pages} {1--40}
  (\bibinfo {year} {2021})}\BibitemShut {NoStop}%
\bibitem [{\citenamefont {Laverick}\ \emph {et~al.}(2023)\citenamefont
  {Laverick}, \citenamefont {Warszawski}, \citenamefont {Chantasri},\ and\
  \citenamefont {Wiseman}}]{LWCW23}%
  \BibitemOpen
  \bibfield  {author} {\bibinfo {author} {\bibfnamefont {K.~T.}\ \bibnamefont
  {Laverick}}, \bibinfo {author} {\bibfnamefont {P.}~\bibnamefont
  {Warszawski}}, \bibinfo {author} {\bibfnamefont {A.}~\bibnamefont
  {Chantasri}}, \ and\ \bibinfo {author} {\bibfnamefont {H.~M.}\ \bibnamefont
  {Wiseman}},\ }\bibfield  {title} {\enquote {\bibinfo {title} {Quantum state
  smoothing cannot be assumed classical even when the filtering and
  retrofiltering are classical},}\ }\href {\doibase
  10.1103/PRXQuantum.4.040340} {\bibfield  {journal} {\bibinfo  {journal} {PRX
  Quantum}\ }\textbf {\bibinfo {volume} {4}},\ \bibinfo {pages} {040340}
  (\bibinfo {year} {2023})}\BibitemShut {NoStop}%
\bibitem [{\citenamefont {Watanabe}(1955)}]{Watanabe55}%
  \BibitemOpen
  \bibfield  {author} {\bibinfo {author} {\bibfnamefont {S.}~\bibnamefont
  {Watanabe}},\ }\bibfield  {title} {\enquote {\bibinfo {title} {{Symmetry of
  physical laws. Part III. Prediction and retrodiction}},}\ }\href {\doibase
  10.1103/RevModPhys.27.179} {\bibfield  {journal} {\bibinfo  {journal} {Rev.
  Mod. Phys.}\ }\textbf {\bibinfo {volume} {27}},\ \bibinfo {pages} {179}
  (\bibinfo {year} {1955})}\BibitemShut {NoStop}%
\bibitem [{\citenamefont {Watanabe}(1956)}]{Watanabe56}%
  \BibitemOpen
  \bibfield  {author} {\bibinfo {author} {\bibfnamefont {S.}~\bibnamefont
  {Watanabe}},\ }\bibfield  {title} {\enquote {\bibinfo {title} {{Symmetry in
  time and Tanikawa's Method of superquantization in regard to negative energy
  fields}},}\ }\href {\doibase 10.1143/ptp.15.523} {\bibfield  {journal}
  {\bibinfo  {journal} {Prog. Theor. Phys.}\ }\textbf {\bibinfo {volume}
  {15}},\ \bibinfo {pages} {523--535} (\bibinfo {year} {1956})}\BibitemShut
  {NoStop}%
\bibitem [{\citenamefont {Aharonov}\ \emph {et~al.}(1964)\citenamefont
  {Aharonov}, \citenamefont {Bergmann},\ and\ \citenamefont
  {Lebowitz}}]{ABL64}%
  \BibitemOpen
  \bibfield  {author} {\bibinfo {author} {\bibfnamefont {Y.}~\bibnamefont
  {Aharonov}}, \bibinfo {author} {\bibfnamefont {P.~G.}\ \bibnamefont
  {Bergmann}}, \ and\ \bibinfo {author} {\bibfnamefont {J.~L.}\ \bibnamefont
  {Lebowitz}},\ }\bibfield  {title} {\enquote {\bibinfo {title} {{Time symmetry
  in the Quantum process of measurement}},}\ }\href {\doibase
  10.1103/PhysRev.134.B1410} {\bibfield  {journal} {\bibinfo  {journal} {Phys.
  Rev.}\ }\textbf {\bibinfo {volume} {134}},\ \bibinfo {pages} {B1410--B1416}
  (\bibinfo {year} {1964})}\BibitemShut {NoStop}%
\bibitem [{\citenamefont {Laverick}\ \emph
  {et~al.}(2021{\natexlab{a}})\citenamefont {Laverick}, \citenamefont
  {Chantasri},\ and\ \citenamefont {Wiseman}}]{LCW-PRA21}%
  \BibitemOpen
  \bibfield  {author} {\bibinfo {author} {\bibfnamefont {K.~T.}\ \bibnamefont
  {Laverick}}, \bibinfo {author} {\bibfnamefont {A.}~\bibnamefont {Chantasri}},
  \ and\ \bibinfo {author} {\bibfnamefont {H.~M.}\ \bibnamefont {Wiseman}},\
  }\bibfield  {title} {\enquote {\bibinfo {title} {Linear gaussian quantum
  state smoothing: Understanding the optimal unravelings for {A}lice to
  estimate {B}ob's state},}\ }\href {\doibase 10.1103/PhysRevA.103.012213}
  {\bibfield  {journal} {\bibinfo  {journal} {Phys. Rev. A}\ }\textbf {\bibinfo
  {volume} {103}},\ \bibinfo {pages} {012213} (\bibinfo {year}
  {2021}{\natexlab{a}})}\BibitemShut {NoStop}%
\bibitem [{\citenamefont {Aharonov}\ \emph {et~al.}(1988)\citenamefont
  {Aharonov}, \citenamefont {Albert},\ and\ \citenamefont {Vaidman}}]{AAV88}%
  \BibitemOpen
  \bibfield  {author} {\bibinfo {author} {\bibfnamefont {Y.}~\bibnamefont
  {Aharonov}}, \bibinfo {author} {\bibfnamefont {D.~Z.}\ \bibnamefont
  {Albert}}, \ and\ \bibinfo {author} {\bibfnamefont {L.}~\bibnamefont
  {Vaidman}},\ }\bibfield  {title} {\enquote {\bibinfo {title} {{How the result
  of a measurement of a component of the spin of a spin-1/2 particle can turn
  out to be 100}},}\ }\href {\doibase 10.1103/PhysRevLett.60.1351} {\bibfield
  {journal} {\bibinfo  {journal} {Phys. Rev. Lett.}\ }\textbf {\bibinfo
  {volume} {60}},\ \bibinfo {pages} {1351--1354} (\bibinfo {year}
  {1988})}\BibitemShut {NoStop}%
\bibitem [{\citenamefont {Belavkin}(1999)}]{Bel99}%
  \BibitemOpen
  \bibfield  {author} {\bibinfo {author} {\bibfnamefont {V.~P.}\ \bibnamefont
  {Belavkin}},\ }\bibfield  {title} {\enquote {\bibinfo {title} {{Measurement,
  filtering and control in quantum open dynamical systems}},}\ }\href {\doibase
  doi:10.1016/S0034-4877(00)86386-7} {\bibfield  {journal} {\bibinfo  {journal}
  {Rep. Math. Phys.}\ }\textbf {\bibinfo {volume} {43}},\ \bibinfo {pages}
  {A405} (\bibinfo {year} {1999})}\BibitemShut {NoStop}%
\bibitem [{\citenamefont {Carmichael}(1993)}]{Carmichael93}%
  \BibitemOpen
  \bibfield  {author} {\bibinfo {author} {\bibfnamefont {H.~J.}\ \bibnamefont
  {Carmichael}},\ }\href@noop {} {\emph {\bibinfo {title} {{An open systems
  approach to quantum optics}}}}\ (\bibinfo  {publisher} {Springer},\ \bibinfo
  {address} {Berlin},\ \bibinfo {year} {1993})\BibitemShut {NoStop}%
\bibitem [{\citenamefont {Wiseman}\ and\ \citenamefont
  {Milburn}(2010)}]{WisMil10}%
  \BibitemOpen
  \bibfield  {author} {\bibinfo {author} {\bibfnamefont {H.~M.}\ \bibnamefont
  {Wiseman}}\ and\ \bibinfo {author} {\bibfnamefont {G.~J.}\ \bibnamefont
  {Milburn}},\ }\href@noop {} {\emph {\bibinfo {title} {{Quantum measurement
  and control}}}}\ (\bibinfo  {publisher} {Cambridge University Press},\
  \bibinfo {address} {Cambridge, England},\ \bibinfo {year} {2010})\BibitemShut
  {NoStop}%
\bibitem [{\citenamefont {Laverick}\ \emph
  {et~al.}(2021{\natexlab{b}})\citenamefont {Laverick}, \citenamefont
  {Guevara},\ and\ \citenamefont {Wiseman}}]{LGW-PRA21}%
  \BibitemOpen
  \bibfield  {author} {\bibinfo {author} {\bibfnamefont {K.~T.}\ \bibnamefont
  {Laverick}}, \bibinfo {author} {\bibfnamefont {I.}~\bibnamefont {Guevara}}, \
  and\ \bibinfo {author} {\bibfnamefont {H.~M.}\ \bibnamefont {Wiseman}},\
  }\bibfield  {title} {\enquote {\bibinfo {title} {{Quantum state smoothing as
  an optimal Bayesian estimation problem with three different cost
  functions}},}\ }\href {\doibase 10.1103/PhysRevA.104.032213} {\bibfield
  {journal} {\bibinfo  {journal} {Phys. Rev. A}\ }\textbf {\bibinfo {volume}
  {104}},\ \bibinfo {pages} {032213} (\bibinfo {year}
  {2021}{\natexlab{b}})}\BibitemShut {NoStop}%
\bibitem [{\citenamefont {Chantasri}\ \emph {et~al.}(2019)\citenamefont
  {Chantasri}, \citenamefont {Guevara},\ and\ \citenamefont {Wiseman}}]{CGW19}%
  \BibitemOpen
  \bibfield  {author} {\bibinfo {author} {\bibfnamefont {A.}~\bibnamefont
  {Chantasri}}, \bibinfo {author} {\bibfnamefont {I.}~\bibnamefont {Guevara}},
  \ and\ \bibinfo {author} {\bibfnamefont {H.~M.}\ \bibnamefont {Wiseman}},\
  }\bibfield  {title} {\enquote {\bibinfo {title} {Quantum state smoothing: why
  the types of observed and unobserved measurements matter},}\ }\href {\doibase
  10.1088/1367-2630/ab396e} {\bibfield  {journal} {\bibinfo  {journal} {New J.
  Phys.}\ }\textbf {\bibinfo {volume} {21}},\ \bibinfo {pages} {083039}
  (\bibinfo {year} {2019})}\BibitemShut {NoStop}%
\bibitem [{\citenamefont {Laverick}\ \emph {et~al.}(2019)\citenamefont
  {Laverick}, \citenamefont {Chantasri},\ and\ \citenamefont
  {Wiseman}}]{LCW19}%
  \BibitemOpen
  \bibfield  {author} {\bibinfo {author} {\bibfnamefont {K.~T.}\ \bibnamefont
  {Laverick}}, \bibinfo {author} {\bibfnamefont {A.}~\bibnamefont {Chantasri}},
  \ and\ \bibinfo {author} {\bibfnamefont {H.~M.}\ \bibnamefont {Wiseman}},\
  }\bibfield  {title} {\enquote {\bibinfo {title} {{Quantum state smoothing for
  linear Gaussian systems}},}\ }\href {\doibase 10.1103/PhysRevLett.122.190402}
  {\bibfield  {journal} {\bibinfo  {journal} {Phys. Rev. Lett.}\ }\textbf
  {\bibinfo {volume} {122}},\ \bibinfo {pages} {190402} (\bibinfo {year}
  {2019})}\BibitemShut {NoStop}%
\bibitem [{\citenamefont {Petz}(1986)}]{petz1986}%
  \BibitemOpen
  \bibfield  {author} {\bibinfo {author} {\bibfnamefont {D.}~\bibnamefont
  {Petz}},\ }\bibfield  {title} {\enquote {\bibinfo {title} {Sufficient
  subalgebras and the relative entropy of states of a von neumann algebra},}\
  }\href {\doibase 10.1007/BF01212345} {\bibfield  {journal} {\bibinfo
  {journal} {Commun. Math. Phys.}\ }\textbf {\bibinfo {volume} {105}},\
  \bibinfo {pages} {123--131} (\bibinfo {year} {1986})}\BibitemShut {NoStop}%
\bibitem [{\citenamefont {Petz}(1988)}]{petz1988}%
  \BibitemOpen
  \bibfield  {author} {\bibinfo {author} {\bibfnamefont {D.}~\bibnamefont
  {Petz}},\ }\bibfield  {title} {\enquote {\bibinfo {title} {Sufficiency of
  channels over von {N}eumann algebra},}\ }\href {\doibase
  10.1093/qmath/39.1.97} {\bibfield  {journal} {\bibinfo  {journal} {Q. J.
  Math.}\ }\textbf {\bibinfo {volume} {39}},\ \bibinfo {pages} {97--108}
  (\bibinfo {year} {1988})}\BibitemShut {NoStop}%
\bibitem [{\citenamefont {Fuchs}(2003)}]{Fuchs03}%
  \BibitemOpen
  \bibfield  {author} {\bibinfo {author} {\bibfnamefont {C.~A.}\ \bibnamefont
  {Fuchs}},\ }\bibfield  {title} {\enquote {\bibinfo {title} {Quantum mechanics
  as quantum information, mostly},}\ }\href {\doibase
  10.1080/09500340308234548} {\bibfield  {journal} {\bibinfo  {journal} {J.
  Mod. Opt.}\ }\textbf {\bibinfo {volume} {50}},\ \bibinfo {pages} {987--1023}
  (\bibinfo {year} {2003})}\BibitemShut {NoStop}%
\bibitem [{\citenamefont {Laverick}\ \emph
  {et~al.}(2021{\natexlab{c}})\citenamefont {Laverick}, \citenamefont
  {Chantasri},\ and\ \citenamefont {Wiseman}}]{LCW-QS21}%
  \BibitemOpen
  \bibfield  {author} {\bibinfo {author} {\bibfnamefont {K.~T.}\ \bibnamefont
  {Laverick}}, \bibinfo {author} {\bibfnamefont {A.}~\bibnamefont {Chantasri}},
  \ and\ \bibinfo {author} {\bibfnamefont {H.~M.}\ \bibnamefont {Wiseman}},\
  }\bibfield  {title} {\enquote {\bibinfo {title} {{General criteria for
  quantum state smoothing with necessary and sufficient criteria for linear
  Gaussian quantum systems}},}\ }\href {\doibase 10.1007/s40509-020-00225-7}
  {\bibfield  {journal} {\bibinfo  {journal} {Quantum Stud.: Math. Found.}\
  }\textbf {\bibinfo {volume} {8}},\ \bibinfo {pages} {37--50} (\bibinfo {year}
  {2021}{\natexlab{c}})}\BibitemShut {NoStop}%
\bibitem [{\citenamefont {Walls}\ and\ \citenamefont
  {Milburn}(1994)}]{WallsMilb94}%
  \BibitemOpen
  \bibfield  {author} {\bibinfo {author} {\bibfnamefont {D.~F.}\ \bibnamefont
  {Walls}}\ and\ \bibinfo {author} {\bibfnamefont {G.~J.}\ \bibnamefont
  {Milburn}},\ }\href@noop {} {\emph {\bibinfo {title} {Quantum Optics}}}\
  (\bibinfo  {publisher} {Springer},\ \bibinfo {address} {Berlin},\ \bibinfo
  {year} {1994})\BibitemShut {NoStop}%
\bibitem [{\citenamefont {Gardiner}(2004)}]{GardinerBook}%
  \BibitemOpen
  \bibfield  {author} {\bibinfo {author} {\bibfnamefont {C.~W.}\ \bibnamefont
  {Gardiner}},\ }\href@noop {} {\emph {\bibinfo {title} {Handbook of stochastic
  methods for physics, chemistry and the natural sciences}}},\ \bibinfo
  {edition} {3rd}\ ed.\ (\bibinfo  {publisher} {Springer},\ \bibinfo {address}
  {Berlin},\ \bibinfo {year} {2004})\BibitemShut {NoStop}%
\bibitem [{\citenamefont {Speyer}\ and\ \citenamefont
  {Chung}(2008)}]{SpeChu08}%
  \BibitemOpen
  \bibfield  {author} {\bibinfo {author} {\bibfnamefont {J.~L.}\ \bibnamefont
  {Speyer}}\ and\ \bibinfo {author} {\bibfnamefont {W.~H.}\ \bibnamefont
  {Chung}},\ }\href@noop {} {\emph {\bibinfo {title} {Stochastic Processes,
  Estimation, and Control}}}\ (\bibinfo  {publisher} {SIAM},\ \bibinfo
  {address} {Philadelphia},\ \bibinfo {year} {2008})\BibitemShut {NoStop}%
\bibitem [{SM()}]{SM}%
  \BibitemOpen
  \href@noop {} {}\bibinfo {note} {See the Supplementary Material for
  details.}\BibitemShut {Stop}%
\bibitem [{\citenamefont {Buscemi}\ and\ \citenamefont
  {Scarani}(2021)}]{BusSca21}%
  \BibitemOpen
  \bibfield  {author} {\bibinfo {author} {\bibfnamefont {Francesco}\
  \bibnamefont {Buscemi}}\ and\ \bibinfo {author} {\bibfnamefont {Valerio}\
  \bibnamefont {Scarani}},\ }\bibfield  {title} {\enquote {\bibinfo {title}
  {Fluctuation theorems from bayesian retrodiction},}\ }\href {\doibase
  10.1103/PhysRevE.103.052111} {\bibfield  {journal} {\bibinfo  {journal}
  {Phys. Rev. E}\ }\textbf {\bibinfo {volume} {103}},\ \bibinfo {pages}
  {052111} (\bibinfo {year} {2021})}\BibitemShut {NoStop}%
\bibitem [{\citenamefont {Breuer}\ and\ \citenamefont
  {Petruccione}(2006)}]{BrePet06}%
  \BibitemOpen
  \bibfield  {author} {\bibinfo {author} {\bibfnamefont {H.-P.}\ \bibnamefont
  {Breuer}}\ and\ \bibinfo {author} {\bibfnamefont {F.}~\bibnamefont
  {Petruccione}},\ }\href@noop {} {\emph {\bibinfo {title} {{The theory of open
  quantum systems}}}}\ (\bibinfo  {publisher} {Oxford University Press},\
  \bibinfo {address} {New York},\ \bibinfo {year} {2006})\BibitemShut {NoStop}%
\bibitem [{\citenamefont {Manikandan}\ and\ \citenamefont
  {Jordan}(2019)}]{ManJor19}%
  \BibitemOpen
  \bibfield  {author} {\bibinfo {author} {\bibfnamefont {S.~K.}\ \bibnamefont
  {Manikandan}}\ and\ \bibinfo {author} {\bibfnamefont {A.~N.}\ \bibnamefont
  {Jordan}},\ }\bibfield  {title} {\enquote {\bibinfo {title} {Time reversal
  symmetry of generalized quantum measurements with past and future boundary
  conditions},}\ }\href {\doibase 10.1007/s40509-019-00182-w} {\bibfield
  {journal} {\bibinfo  {journal} {Quantum Stud.: Math. Found.}\ }\textbf
  {\bibinfo {volume} {6}},\ \bibinfo {pages} {241--268} (\bibinfo {year}
  {2019})}\BibitemShut {NoStop}%
\bibitem [{\citenamefont {Manikandan}\ \emph {et~al.}(2019)\citenamefont
  {Manikandan}, \citenamefont {Elouard},\ and\ \citenamefont
  {Jordan}}]{EloJor19}%
  \BibitemOpen
  \bibfield  {author} {\bibinfo {author} {\bibfnamefont {S.~K.}\ \bibnamefont
  {Manikandan}}, \bibinfo {author} {\bibfnamefont {C.}~\bibnamefont {Elouard}},
  \ and\ \bibinfo {author} {\bibfnamefont {A.~N.}\ \bibnamefont {Jordan}},\
  }\bibfield  {title} {\enquote {\bibinfo {title} {Fluctuation theorems for
  continuous quantum measurements and absolute irreversibility},}\ }\href
  {\doibase 10.1103/PhysRevA.99.022117} {\bibfield  {journal} {\bibinfo
  {journal} {Phys. Rev. A}\ }\textbf {\bibinfo {volume} {99}},\ \bibinfo
  {pages} {022117} (\bibinfo {year} {2019})}\BibitemShut {NoStop}%
\bibitem [{\citenamefont {Surace}\ and\ \citenamefont {Scandi}(2023)}]{SS23}%
  \BibitemOpen
  \bibfield  {author} {\bibinfo {author} {\bibfnamefont {J.}~\bibnamefont
  {Surace}}\ and\ \bibinfo {author} {\bibfnamefont {M.}~\bibnamefont
  {Scandi}},\ }\bibfield  {title} {\enquote {\bibinfo {title} {State retrieval
  beyond {B}ayes' retrodiction},}\ }\href {\doibase 10.22331/q-2023-04-27-990}
  {\bibfield  {journal} {\bibinfo  {journal} {{Quantum}}\ }\textbf {\bibinfo
  {volume} {7}},\ \bibinfo {pages} {990} (\bibinfo {year} {2023})}\BibitemShut
  {NoStop}%
\bibitem [{\citenamefont {Parzygnat}\ and\ \citenamefont
  {Fullwood}(2023)}]{PF23}%
  \BibitemOpen
  \bibfield  {author} {\bibinfo {author} {\bibfnamefont {A.~J.}\ \bibnamefont
  {Parzygnat}}\ and\ \bibinfo {author} {\bibfnamefont {J.}~\bibnamefont
  {Fullwood}},\ }\bibfield  {title} {\enquote {\bibinfo {title} {From
  time-reversal symmetry to quantum bayes' rules},}\ }\href {\doibase
  10.1103/PRXQuantum.4.020334} {\bibfield  {journal} {\bibinfo  {journal} {PRX
  Quantum}\ }\textbf {\bibinfo {volume} {4}},\ \bibinfo {pages} {020334}
  (\bibinfo {year} {2023})}\BibitemShut {NoStop}%
\bibitem [{\citenamefont {Cenxin}\ \emph {et~al.}(2023)\citenamefont {Cenxin},
  \citenamefont {Onggadinata}, \citenamefont {Kaszlikowski},\ and\
  \citenamefont {Scarani}}]{CS23}%
  \BibitemOpen
  \bibfield  {author} {\bibinfo {author} {\bibfnamefont {A.~C.}\ \bibnamefont
  {Cenxin}}, \bibinfo {author} {\bibfnamefont {K.}~\bibnamefont {Onggadinata}},
  \bibinfo {author} {\bibfnamefont {D.}~\bibnamefont {Kaszlikowski}}, \ and\
  \bibinfo {author} {\bibfnamefont {V.}~\bibnamefont {Scarani}},\ }\bibfield
  {title} {\enquote {\bibinfo {title} {Quantum {B}ayesian inference in
  quasiprobability representations},}\ }\href {\doibase
  10.1103/PRXQuantum.4.020352} {\bibfield  {journal} {\bibinfo  {journal} {PRX
  Quantum}\ }\textbf {\bibinfo {volume} {4}},\ \bibinfo {pages} {020352}
  (\bibinfo {year} {2023})}\BibitemShut {NoStop}%
\bibitem [{\citenamefont {Parzygnat}\ and\ \citenamefont
  {Buscemi}(2023)}]{PB23}%
  \BibitemOpen
  \bibfield  {author} {\bibinfo {author} {\bibfnamefont {A.~J.}\ \bibnamefont
  {Parzygnat}}\ and\ \bibinfo {author} {\bibfnamefont {F.}~\bibnamefont
  {Buscemi}},\ }\bibfield  {title} {\enquote {\bibinfo {title} {Axioms for
  retrodiction: achieving time-reversal symmetry with a prior},}\ }\href
  {\doibase 10.22331/q-2023-05-23-1013} {\bibfield  {journal} {\bibinfo
  {journal} {{Quantum}}\ }\textbf {\bibinfo {volume} {7}},\ \bibinfo {pages}
  {1013} (\bibinfo {year} {2023})}\BibitemShut {NoStop}%
\bibitem [{\citenamefont {Aw}\ \emph {et~al.}(2024)\citenamefont {Aw},
  \citenamefont {Zaw}, \citenamefont {Balanz\'o-Juand\'o},\ and\ \citenamefont
  {Scarani}}]{CLS24}%
  \BibitemOpen
  \bibfield  {author} {\bibinfo {author} {\bibfnamefont {C.~C.}\ \bibnamefont
  {Aw}}, \bibinfo {author} {\bibfnamefont {L.~H.}\ \bibnamefont {Zaw}},
  \bibinfo {author} {\bibfnamefont {M.}~\bibnamefont {Balanz\'o-Juand\'o}}, \
  and\ \bibinfo {author} {\bibfnamefont {V.}~\bibnamefont {Scarani}},\
  }\bibfield  {title} {\enquote {\bibinfo {title} {Role of dilations in
  reversing physical processes: Tabletop reversibility and generalized thermal
  operations},}\ }\href {\doibase 10.1103/PRXQuantum.5.010332} {\bibfield
  {journal} {\bibinfo  {journal} {PRX Quantum}\ }\textbf {\bibinfo {volume}
  {5}},\ \bibinfo {pages} {010332} (\bibinfo {year} {2024})}\BibitemShut
  {NoStop}%
\bibitem [{\citenamefont {Barnum}\ and\ \citenamefont {Knill}(2002)}]{BK02}%
  \BibitemOpen
  \bibfield  {author} {\bibinfo {author} {\bibfnamefont {H.}~\bibnamefont
  {Barnum}}\ and\ \bibinfo {author} {\bibfnamefont {E.}~\bibnamefont {Knill}},\
  }\bibfield  {title} {\enquote {\bibinfo {title} {Reversing quantum dynamics
  with near-optimal quantum and classical fidelity},}\ }\href {\doibase
  10.1063/1.1459754} {\bibfield  {journal} {\bibinfo  {journal} {J. Math.
  Phys.}\ }\textbf {\bibinfo {volume} {43}},\ \bibinfo {pages} {2097--2106}
  (\bibinfo {year} {2002})}\BibitemShut {NoStop}%
\bibitem [{\citenamefont {Ng}\ and\ \citenamefont {Mandayam}(2010)}]{NHK10}%
  \BibitemOpen
  \bibfield  {author} {\bibinfo {author} {\bibfnamefont {H.~K.}\ \bibnamefont
  {Ng}}\ and\ \bibinfo {author} {\bibfnamefont {P.}~\bibnamefont {Mandayam}},\
  }\bibfield  {title} {\enquote {\bibinfo {title} {Simple approach to
  approximate quantum error correction based on the transpose channel},}\
  }\href {\doibase 10.1103/PhysRevA.81.062342} {\bibfield  {journal} {\bibinfo
  {journal} {Phys. Rev. A}\ }\textbf {\bibinfo {volume} {81}},\ \bibinfo
  {pages} {062342} (\bibinfo {year} {2010})}\BibitemShut {NoStop}%
\bibitem [{\citenamefont {Kwon}\ \emph {et~al.}(2022)\citenamefont {Kwon},
  \citenamefont {Mukherjee},\ and\ \citenamefont {Kim}}]{KMK22}%
  \BibitemOpen
  \bibfield  {author} {\bibinfo {author} {\bibfnamefont {H.}~\bibnamefont
  {Kwon}}, \bibinfo {author} {\bibfnamefont {R.}~\bibnamefont {Mukherjee}}, \
  and\ \bibinfo {author} {\bibfnamefont {M.~S.}\ \bibnamefont {Kim}},\
  }\bibfield  {title} {\enquote {\bibinfo {title} {Reversing lindblad dynamics
  via continuous petz recovery map},}\ }\href {\doibase
  10.1103/PhysRevLett.128.020403} {\bibfield  {journal} {\bibinfo  {journal}
  {Phys. Rev. Lett.}\ }\textbf {\bibinfo {volume} {128}},\ \bibinfo {pages}
  {020403} (\bibinfo {year} {2022})}\BibitemShut {NoStop}%
\bibitem [{\citenamefont {Braunstein}\ and\ \citenamefont {van
  Loock}(2005)}]{Bra05}%
  \BibitemOpen
  \bibfield  {author} {\bibinfo {author} {\bibfnamefont {S.~L.}\ \bibnamefont
  {Braunstein}}\ and\ \bibinfo {author} {\bibfnamefont {P.}~\bibnamefont {van
  Loock}},\ }\bibfield  {title} {\enquote {\bibinfo {title} {Quantum
  information with continuous variables},}\ }\href {\doibase
  10.1103/RevModPhys.77.513} {\bibfield  {journal} {\bibinfo  {journal} {Rev.
  Mod. Phys.}\ }\textbf {\bibinfo {volume} {77}},\ \bibinfo {pages} {513--577}
  (\bibinfo {year} {2005})}\BibitemShut {NoStop}%
\bibitem [{\citenamefont {Bowen}\ and\ \citenamefont
  {Milburn}(2016)}]{BowMil16}%
  \BibitemOpen
  \bibfield  {author} {\bibinfo {author} {\bibfnamefont {W.~P.}\ \bibnamefont
  {Bowen}}\ and\ \bibinfo {author} {\bibfnamefont {G.~J.}\ \bibnamefont
  {Milburn}},\ }\href@noop {} {\emph {\bibinfo {title} {Quantum
  Optomechanics}}}\ (\bibinfo  {publisher} {CRC Press},\ \bibinfo {address}
  {Boca Raton},\ \bibinfo {year} {2016})\BibitemShut {NoStop}%
\bibitem [{\citenamefont {Laverick}(2021)}]{Laverick21}%
  \BibitemOpen
  \bibfield  {author} {\bibinfo {author} {\bibfnamefont {K.~T.}\ \bibnamefont
  {Laverick}},\ }\bibfield  {title} {\enquote {\bibinfo {title} {Quantum
  rauch-tung-striebel smoothed state},}\ }\href {\doibase
  10.1103/PhysRevResearch.3.033196} {\bibfield  {journal} {\bibinfo  {journal}
  {Phys. Rev. Res.}\ }\textbf {\bibinfo {volume} {3}},\ \bibinfo {pages}
  {033196} (\bibinfo {year} {2021})}\BibitemShut {NoStop}%
\bibitem [{\citenamefont {Barnett}\ \emph {et~al.}(2000)\citenamefont
  {Barnett}, \citenamefont {Pegg},\ and\ \citenamefont
  {Jeffers}}]{Barnett2000}%
  \BibitemOpen
  \bibfield  {author} {\bibinfo {author} {\bibfnamefont {S.~M.}\ \bibnamefont
  {Barnett}}, \bibinfo {author} {\bibfnamefont {D.~T.}\ \bibnamefont {Pegg}}, \
  and\ \bibinfo {author} {\bibfnamefont {J.}~\bibnamefont {Jeffers}},\
  }\bibfield  {title} {\enquote {\bibinfo {title} {Bayes' theorem and quantum
  retrodiction},}\ }\href {\doibase 10.1080/09500340008232431} {\bibfield
  {journal} {\bibinfo  {journal} {J. Mod. Opt.}\ }\textbf {\bibinfo {volume}
  {47}},\ \bibinfo {pages} {1779--1789} (\bibinfo {year} {2000})}\BibitemShut
  {NoStop}%
\bibitem [{\citenamefont {Leifer}\ and\ \citenamefont {Spekkens}(2013)}]{LS13}%
  \BibitemOpen
  \bibfield  {author} {\bibinfo {author} {\bibfnamefont {M.~S.}\ \bibnamefont
  {Leifer}}\ and\ \bibinfo {author} {\bibfnamefont {R.~W.}\ \bibnamefont
  {Spekkens}},\ }\bibfield  {title} {\enquote {\bibinfo {title} {Towards a
  formulation of quantum theory as a causally neutral theory of bayesian
  inference},}\ }\href {\doibase 10.1103/PhysRevA.88.052130} {\bibfield
  {journal} {\bibinfo  {journal} {Phys. Rev. A}\ }\textbf {\bibinfo {volume}
  {88}},\ \bibinfo {pages} {052130} (\bibinfo {year} {2013})}\BibitemShut
  {NoStop}%
\bibitem [{\citenamefont {Guevara}\ and\ \citenamefont
  {Wiseman}(2020)}]{GueWis20}%
  \BibitemOpen
  \bibfield  {author} {\bibinfo {author} {\bibfnamefont {I.}~\bibnamefont
  {Guevara}}\ and\ \bibinfo {author} {\bibfnamefont {H.~M.}\ \bibnamefont
  {Wiseman}},\ }\bibfield  {title} {\enquote {\bibinfo {title} {Completely
  positive quantum trajectories with applications to quantum state
  smoothing},}\ }\href {\doibase 10.1103/PhysRevA.102.052217} {\bibfield
  {journal} {\bibinfo  {journal} {Phys. Rev. A}\ }\textbf {\bibinfo {volume}
  {102}},\ \bibinfo {pages} {052217} (\bibinfo {year} {2020})}\BibitemShut
  {NoStop}%
\bibitem [{\citenamefont {Kraus}(1983)}]{Kraus83}%
  \BibitemOpen
  \bibfield  {author} {\bibinfo {author} {\bibfnamefont {K.}~\bibnamefont
  {Kraus}},\ }\href@noop {} {\emph {\bibinfo {title} {States, Effects, and
  Operations: Fundamental Notions of Quantum Theory}}},\ \bibinfo {series}
  {Lecture notes in Physics}, Vol.\ \bibinfo {volume} {190}\ (\bibinfo
  {publisher} {Springer},\ \bibinfo {address} {Berlin},\ \bibinfo {year}
  {1983})\BibitemShut {NoStop}%
\end{thebibliography}
\end{document}